\documentclass[aps,prc,amsfonts,superscriptaddress,
twocolumn,
byrevtex
]{revtex4}

\usepackage{epsfig}
\usepackage{amsmath}
\usepackage{amssymb}
\usepackage{amsfonts}
\usepackage{color}

\newcommand{\JMcomm}[1]{{\textcolor{black}{ #1}}}

\newcommand{\old}[1]{{\textcolor{black}{ }}}

\renewcommand{\vec}[1]{\mathbf{#1}}

\begin{document}

\title{An alternate well-founded way to treat the center-of-mass correlations:\\
use of a local center-of-mass correlations potential.}

\author{J{\'e}r{\'e}mie Messud}
%
\affiliation{Universit{\'e} Bordeaux, CNRS/IN2P3,
             Centre d'Etudes Nucl{\'e}aires de Bordeaux Gradignan, UMR5797,
             F-33175 Gradignan, France.\\
             Now at: CGGVeritas, F-91341 Massy, France.
}
%
\date{\today} 
\begin{abstract}
The recently developed ``internal" Density Functional Theory provides an existence theorem for a
local potential that contains the center-of-mass correlations effects.
The knowledge of the corresponding energy functional would provide a much cheaper way than 
projection techniques to treat these correlations.
The aim of this article is to construct such a functional.
We propose a well-founded method, suitable for Fermions as well as for Bosons,
which does not require any free parameters.
\end{abstract}


\maketitle 

%
%
\section{Introduction.}

One of the most obvious symmetries of isolated self-bound systems
(such as atomic nuclei, helium droplets, or molecular systems where nuclei are treated explicitly)
is translational invariance.
Translational invariance of self-bound Hamiltonians ensures Galilean invariance 
of the wave function, so that the center-of-mass (c.m.) properties can be separated from the
``internal" properties (that are of experimental interest).
As a consequence, one laboratory coordinate is redundant 
for the description of the internal properties,
that produces c.m. correlations.

A numerically manageable and succesful way to describe self-bound systems is to use
mean-field-like calculations with effective interactions.
The corresponding equations are often justified starting from the Hartree-Fock (HF) framework,
which sacrifices by construction ``Galilean invariance for the sake of the Pauli principle", 
to quote Ref.~\cite{Schm01a}.
As a consequence, the c.m. correlations
are treated incorrectly
(in an equivalent manner, the redundant coordinate problem is treated incorrectly).
This introduces a spurious coupling between the internal properties and the c.m. motion in an HF framework
that affects the energy and other observables~\cite{RS80}.

A way to overcome this problem in the stationary case is to perform projected HF,
where projection before variation on c.m.\ momentum restores
translational or Galilean invariance of the wave-function.
Peierls and Yoccoz proposed a single projection method to restore translational invariance \cite{Pei57}.
Later, Peierls and Thouless proposed a double projection method to restore the more fundamental Galilean invariance \cite{Pei62}.
To our knowledge, all numerical calculations that treat the c.m. correlations by projection before variation have
been done using the Peierls and Yoccoz method \cite{Mar83,Mar83-2,Sch90,Dob09}, thus not restoring the full Galilean invariance.
Moreover, the price to pay is the abandon of the independent-particle 
description and a large numerical cost \cite{Schm01a,Rod04a,Pei62,Ben03}.
Indeed, projection techniques require ``an order of magnitude more computing time than
the underlying mean-field-like calculations", to quote Ref.~\cite{Sch90},
which is prejudicial for the description of intermediate-sized systems.
This led to the development of various approximate methods to treat the c.m. correlations; see Ref.~\cite{ben00} for an overview.
\JMcomm{
For instance, a common method is to add a $-<\frac{\mathbf{P}^2}{2mN}>$ term in the energy functional
(more details will be given in \S\ref{sec:current_cm}).
}But the success of those methods is not systematic and the approximations done not completely justified.

In the time-dependent case, the spurious c.m.\ motion problem remains~\cite{Irv80,Uma09},
but the situation is trickier as the projected HF method becomes unmanageable even for very small self-bound systems~\cite{Uma09}.
It thus remains an open problem to develop a rigorous and numerically inexpensive scheme to treat the c.m. correlations,
which \JMcomm{would go beyond standard approximations and} remain usable in the time-dependent case.

The search for such a scheme has not yet been pursued extensively, perhaps because it is sometimes
thought that the c.m. correlations problem concerns only very small self-bound systems.
But the c.m correlations can have a non negligible effect even for intermediate-sized systems.
For instance, it has been shown that
c.m. correlations are non-negligible for all nuclei heavier than $^{16}$O \cite{Sch01,Mar83-2,Sch90}.
This reinforces the necessity to develop a numerically  manageable method to treat them.

A rigorous alternative and \textit{a priori} numerically much less costly way to take into account those correlations
has been revealed by the recently developed ``internal" Density Functional Theory (DFT) and Kohn-Sham (KS) scheme \cite{Mes09-1,Mes09-2,Mes11}.
Differing from standard DFT~\cite{Hoh64,Koh65,Koh99,Dre90}, it is formulated in the c.m frame of a self-bound system and
proves that the c.m.\ correlations can be included in the energy functional and thus in a local KS potential \cite{Mes09-1,Mes09-2,Mes11}.
In addition to the fact that it gives a much more fundamental justification than the HF framework
to the use of mean-field-like calculations with effective interactions for the description of self-bound systems,
it shows that there would be no need for a c.m.\ projection if the ultimate functional were known.
Internal DFT gives an existence theorem but not a constructive method.
The aim of the present article is to propose such a constructive method.

The article is organized as follows.
Section II provides a brief review of the internal DFT formalism and underlines the limitations of the commonly used methods to treat the c.m. correlations.
Section III develops a new general form for a local c.m. correlations potential that introduces no free parameter.
Finally, section IV gives convincing numerical results on various model systems.

%
%
\section{Internal DFT and the c.m. correlations functional.}

\subsection{Brief review of the internal DFT formalism.}
\label{sec:intDFTrecall}

We start from a self-bound system composed of $N$ identical particles of mass $m$ and follow the considerations of Ref.~\cite{Mes09-1}.
The coordinates of the particles in any chosen inertial frame of reference (such as
the laboratory) are denoted $\{\vec{r}_i\}$. The c.m. coordinate of the system is denoted:
$$
\vec{R}=\frac{1}{N}\sum_{j=1}^{N}\vec{r}_j
.
$$
The system is described by the following translationally invariant $N$-body Hamiltonian:
\begin{equation}
\label{eq:H}
H 
=   \sum_{i=1}^{N} \frac{\vec{p}^2_i}{2m} 
  + \sum_{\stackrel{i,j=1}{i > j}}^{N} u (\vec{r}_i-\vec{r}_j) 
  + \sum_{i=1}^{N} v^{\text{int}} (\vec{r}_i - \vec{R})
\; ,
\end{equation}
composed of the usual kinetic energy term, a 
2-body potential $u$ which describes the particle-particle
interaction (generalization of the following considerations to 3-body, etc. interactions is straightforward)
and an arbitrary translationally invariant potential $v^{\text{int}}$.
This last potential is an ``internal" potential, i.e., it is defined in the c.m. frame and acts only on the internal properties.
Of course, the potential is zero in the purely isolated self-bound case.
Nevertheless, its form is suitable to model the internal effects of 
fields used in experiments (polarization potentials, etc.)~\cite{Mes09-2}.

We introduce the Jacobi coordinates ${\boldmath{\xi}}_\alpha$
defined as $\mathbf{\xi}_{1} = \mathbf{r}_2-\vec{r}_1$, 
$\mathbf{\xi}_2=\mathbf{r}_3-\frac{\vec{r}_2+\vec{r}_1}{2}$, \ldots,
$\mathbf{\xi}_{N-1} = \frac{N}{N-1} \, (\vec{r}_N - \vec{R})$.
This permits us to separate the Hamiltonian (\ref{eq:H}) into $H = H_{CM} + H_{int}$, where
$H_{CM} = -\hbar^2/(2mN) \Delta_\vec{R}$
is a 1-body Hamiltonian describing the c.m. motion and acting in the $\vec{R}$ space only, and
$H_{int}$ is a $(N-1)$ body-Hamiltonian describing the internal properties and acting in the $\{\xi_\alpha\}$ space only:
\begin{equation}
H_{int}
= \sum_{\alpha=1}^{N-1} \frac{\tau_\alpha^2}{2\mu_\alpha} + U({\boldmath{\xi}}_1, \ldots , {\boldmath{\xi}}_{N-1})
+ V^{\text{int}}({\boldmath{\xi}}_1, \ldots , {\boldmath{\xi}}_{N-1})
.
\nonumber
\end{equation}
$H_{int}$ contains the interaction $u$ and the potential $v^{\text{int}}$, because they can be rewritten as functions of the 
$\{\xi_\alpha\}$ only [denoted respectively $U({\boldmath{\xi}}_1, \ldots , {\boldmath{\xi}}_{N-1})$ 
and $V^{\text{int}}({\boldmath{\xi}}_1, \ldots , {\boldmath{\xi}}_{N-1})$],
and the internal kinetic energy, which is expressed in
terms of the conjugate momentum $\tau_\alpha$ of $\xi_\alpha$ and the reduced masses
$\mu_\alpha = m\frac{\alpha}{\alpha+1}$.
As $[H_{CM},H_{int}]=0$, the eigenstate $\psi$ of $H$ can be written
as a product of the form:
\begin{equation}
\label{eq:psi}
\psi(\vec{r}_1, \ldots , \vec{r}_N) 
= \Gamma(\vec{R}) \; 
  \psi_{int} ({\boldmath{\xi}}_1, \ldots , {\boldmath{\xi}}_{N-1})
,
\;
\end{equation}
where $\Gamma$ and $\psi_{int}$ are defined by the equations:
\begin{eqnarray}
\label{eq:hcm}
-\frac{\hbar^2}{2M} \Delta_\vec{R} \Gamma 
& = & E_{cm} \Gamma \, ,
\\
\label{eq:hint}
H_{int} \psi_{int} 
& = & E_{int} \psi_{int} \, .
\end{eqnarray}
$\Gamma$ is the c.m. wavefunction that describes the motion of the isolated system as 
a whole in any inertial frame of reference.
Since $\Gamma(\vec{R})$ is the solution of the free Schr\"odinger equation, it should be an arbitrary 
stationary plane wave, i.e., infinitely spread and not normalizable.
This leads to the delocalization of $\vec{R}$ and arbitrary c.m.\ energy.
This does not correspond to experimental situations where the system is no longer isolated: interactions with other systems
of the experimental apparatus localize the c.m.
However, this is not a problem since
internal properties that are of experimental interest are fully described by $\psi_{int}$.
Note that $\psi_{int}$ is by definition always normalizable for the ground state of a self-bound system.
The {internal} density associated to $\psi_{int}$ is \cite{Mes09-1,Gir08b,Kaz86}:
%
\begin{eqnarray}
\label{eq:rho_int0}
\lefteqn{ \rho_{int}(\vec{r}) }
\\
& = & N \Big(\frac{N}{N-1}\Big)^3 
      \int \! d\vec{\xi}_1 \cdots  d\mathbf{\xi}_{N-2} \; 
      \big| \psi_{int} \big(\mathbf{\xi}_1, \ldots, \mathbf{\xi}_{N-2},
                   \tfrac{N\vec{r}}{N-1} \big) \big|^2
      \nonumber \\
& = & N \int \! d\vec{r}_1 \cdots d\vec{r}_N \; 
      \delta(\mathbf{R}) |\psi_{int}(\vec{r}_1, \ldots , \vec{r}_N)|^2 \, 
      \delta \big( \vec{r} - (\vec{r}_i-\vec{R}) \big)
.
\nonumber
\end{eqnarray}
$\rho_{int}(\vec{r})$ is normalized to $N$ and $\vec{r}$ is defined in the c.m.\ frame (see the delta relation in the previous equation)
\footnote{
More generally, we can introduce a $\delta(\mathbf{R}-\mathbf{a})$ where
$\mathbf{a}$ is an arbitrary translation vector,
which would lead to perfectly equivalent results.
We chose $\mathbf{a}=\mathbf{0}$ for simplicity, so that the formalism is formulated in the 
c.m. frame.
}.
Note that even if $\psi_{int}$ can be written as a function of the ($N-1$) Jacobi coordinates only,
it can also be written as a function of the $N$ coordinates $\mathbf{r}_i$.
In this case,
one of the coordinates would be redundant \cite{Die96}, which is expressed by the $\delta(\mathbf{R})$ in the previous equation.

The stationary internal DFT theorem demonstrated in various ways in Refs.~\cite{Mes09-1,Eng07,Bar07},
states that for a non-degenerate ground state
and a given kind of particle,
${\psi}_{int}$ can be expressed as a unique
functional of ${\rho}_{int}$, i.e., ${\psi}_{int}[{\rho}_{int}]$.
As a consquence, the ground state internal energy of a self-bound 
system ${E}_{int}=(\psi_{int}[{\rho}_{int}]| H_{int} |\psi_{int}[{\rho}_{int}])$ can also be expressed as a unique
functional of ${\rho}_{int}$.

A practical way to compute ${\rho}_{int}$ is given by the internal KS scheme, developped in Ref.~\cite{Mes09-1}.
To set up this scheme, we assume
that there exists, \textit{in the c.m.\ frame}, a local 
single-particle potential (i.e., a $N$-body non-interacting system)
that can reproduce the exact density $\rho_{int}$ of the interacting system.
We develop $\rho_{int}$ on the corresponding
basis $\{\varphi^i_{int}\}$ of one-body orbitals 
expressed in c.m.\ frame:
\begin{eqnarray}
\rho_{int}(\vec{r}) 
= \sum_{i=1}^N \big| \varphi^i_{int}(\vec{r}) \big|^2
\, .
\end{eqnarray}
We refer the reader to Ref. \cite{Mes11}, \S III.C, for a justification
of the introduction of $N$ orbitals in the KS scheme,
even if only ($N-1$) coordinates are sufficient to describe internal properties.
We implicitly supposed that the particles are Fermions, but a KS scheme to describe Boson condensates can be set up
in a similar manner by 
choosing all the $\varphi^i_{int}$ to be identical.

The KS assumption implies $\varphi^i_{int}[\rho_{int}]$ \cite{Dre90}; thus,
we can rewrite $E_{int}$ as~\cite{Mes09-1}:
\begin{eqnarray}
\label{eq:Eint}
\lefteqn{ E_{int}[\rho_{int}] }
\\
& = & \sum_{i=1}^{N} (\varphi^i_{int}|\frac{\vec{p}^2}{2m}|\varphi^i_{int}) 
      + E_{HXC}[\rho_{int}]
      + \int \! d\vec{r} \; v_{int}(\vec{r}) \, \rho_{int}(\vec{r}) 
\nonumber
,
\end{eqnarray}
where we have introduced
the ``interaction energy functional"
\footnote{
$\gamma_{int}(\vec{r},\vec{r'})$ is the local part of the 2-body \textit{internal} density matrix defined in Ref.~\cite{Mes09-1}, that
is trivially a functional of $\rho_{int}$.
}:
%
%
%
\begin{eqnarray}
\label{eq:XC-}
\lefteqn{ E_{HXC}[\rho_{int}] =}
\\
   && \frac{1}{2} \int \! d\vec{r} \, d\vec{r'} \, 
      \gamma_{int}[\rho_{int}] (\vec{r},\vec{r'}) \, 
      u(\vec{r}-\vec{r'})
      + E_{\Delta kin}[\rho_{int}]
,
\nonumber
\end{eqnarray}
where:
%
%
\begin{eqnarray}
\lefteqn{ E_{\Delta kin}[\rho_{int}] }
\nonumber\\
\label{eq:Exc}
      &=& (\psi_{int}|\sum_{\alpha=1}^{N-1} \frac{\tau_\alpha^2}{2\mu_\alpha}|\psi_{int}) 
      - \sum_{i=1}^{N} (\varphi^i_{int}|\frac{\vec{p}^2}{2m}|\varphi^i_{int}) 
\\
      &=&\int \! d\vec{r}_1 \cdots d\vec{r}_N \delta(\mathbf{R}) \,
      \psi_{int}^*(\vec{r}_1,\dots,\vec{r}_N)
\label{eq:Exc__2}\\
      &&\times \sum_{i=1}^N \frac{\mathbf{p}_i^2}{2m}\psi_{int}(\vec{r}_1,\dots,\vec{r}_N)
      - \sum_{i=1}^{N} (\varphi^i_{int}|\frac{\vec{p}^2}{2m}|\varphi^i_{int}) 
\nonumber.
\end{eqnarray}
%
$E_{HXC}$ traditionally contains the Hartree energy plus the quantum exchange-correlations energy~\cite{Mes09-1}.
We do not explicitly use this decomposition here because common functionals that describe self-bound systems
(such as the Skyrme force in nuclear physics~\cite{RS80}) approximate $E_{HXC}$ as a whole.


We now give particular attention to the $E_{\Delta kin}$ term.
We call the ``interacting" kinetic energy the kinetic energy of the self-bound system, i.e.,
$
\int \! d\vec{r}_1 \cdots d\vec{r}_N \delta(\mathbf{R}) \,
\psi_{int}^*(\vec{r}_1,\dots,\vec{r}_N)\sum_{i=1}^N \frac{\mathbf{p}_i^2}{2m}\psi_{int}(\vec{r}_1,\dots,\vec{r}_N)
$,
and ``non-interacting" kinetic energy the kinetic energy of the KS system, i.e.,
$
\sum_{i=1}^{N} (\varphi^i_{int}|\frac{\vec{p}^2}{2m}|\varphi^i_{int})
$.
We see from Eq.~(\ref{eq:Exc__2}) that $E_{\Delta kin}$ is the difference between those two energies. It thus
contains the exchange and ``standard" correlations
\footnote{
``Standard" correlations mean all the correlations except the c.m. correlations in the following.
\label{foot:1}
} part of the ``interacting"  kinetic energy term,
but also its c.m.\ correlations part (due to $\delta(\mathbf{R})$).
It is the only term of the functional that explicitly contains the c.m. correlations
and represents the main difference with traditional DFT.

Varying $E_{int}[\rho_{int}]$, 
Eq.~(\ref{eq:Eint}), with respect to $\varphi^{i*}_{int}$, and imposing 
orthonormality of the $\{\varphi^i_{int}\}$
leads to ``internal" KS equations:
%
\begin{equation}
\label{eq:varphi_i}
\Big(
- \frac{\hbar^2}{2m}\Delta 
+ U_{HXC}[\rho_{int}] 
+ v_{int}
\Big) \varphi^i_{int} = \epsilon_i \varphi^i_{int}
,
\end{equation}
where $U_{HXC}[\rho_{int}](\vec{r}) 
= \delta E_{HXC}[\rho_{int}] / \delta \rho_{int}(\vec{r})$
is local as expected.
Equations~(\ref{eq:varphi_i}) have the same form as the traditional KS 
equations formulated for non-translationally 
invariant Hamiltonians \cite{Koh65},
but we have justified their use in the c.m.\ frame for
self-bound systems described with translational-invariant Hamiltonians
and shown that the functional form of $U_{HXC}[\rho_{int}] $ differs by the 
inclusion of c.m.\ correlations~\cite{Mes09-1}.

Moreover, we see from Eq.~(\ref{eq:Exc__2}) that
one has to be cautious with the meaning that is given to the non-interacting kinetic energy in
mean-field-like calculations.
Indeed, the non-interacting kinetic energy
cannot be considered as a first order approximation of the interacting kinetic energy
in the general case.
The difference is equal to
$E_{\Delta kin}$,
%
that can be large when c.m. correlations effects are strong,
i.e., for small and intermediate-sized self-bound systems.
For large self-bound systems, $E^{}_{\Delta kin}$ decreases (in relative value) so that the non-interacting and interacting kinetic energies
values approach each other.

Finally, we mention that the internal DFT formalism has been generalized to time-dependent self-bound systems in Ref.~\cite{Mes09-2},
for instance for the description of the collision of two nuclei or laser irradiation,
and multicomponent self-bound systems in Ref.~\cite{Mes11} for the description of self-bound systems composed of different kinds of particles
(atomic nuclei, mixture of $^3$He and $^4$He droplets, and molecular systems where the nuclei are treated explicitly).
This last work permits us to recover the traditional DFT formalism when one kind of particle is much heavier than the others~\cite{Mes11},
\JMcomm{underlining why traditional DFT is well-suited to describe electrons (only) in molecular systems
but not to describe self-bound systems}.

\subsection{\JMcomm{The proposed method to obtain a c.m. correlations functional.}}
\label{sec:method_proposed}

We split the $E_{HXC}$ functional defined by Eq.~(\ref{eq:XC-})
in a more interesting way for our purpose:
\begin{eqnarray}
\label{eq:EdKin2}
E_{HXC}[\rho_{int}] = E_{HXC}^{stand}[\rho_{int}] + E_{cm}[\rho_{int}]
.
\end{eqnarray}
%
%
$E_{HXC}^{stand}$ is the ``standard" many-body interaction energy
(we recall that ``standard" means every interaction energy except that of the c.m. correlations),
that is mostly described by the parametrized functionals 
commonly used for mean-field-like calculations of
self-bound systems (see Refs.~\cite{RS80,Ben03} for a description of functionals used for 
nuclear systems and Ref.~\cite{Bar06} 
for a description of functionals used for helium droplet systems).

$E_{cm}$ is the pure c.m. correlations energy that is
by construction mostly \textit{not} taken into account in commonly used functionals
(except through a renormalization of the mass in the non-interacting kinetic energy term),
which can affect the results; see Ref.~\cite{ben00}.
The goal of this article is to build a
well-founded form for $E_{cm}$ that can be used to describe all self-bound 
systems by simple addition to the commonly used functionals
(\JMcomm{which rigorously implies a refitting of those functionals}),
and is numerically manageable.

The idea is simple: we start from $E_{HXC}$, Eq.~(\ref{eq:XC-}), and neglect all the ``standard" interaction terms.
Then, by definition (\ref{eq:EdKin2}), we are left with $E_{cm}$.
This is equivalent to starting from $E_{\Delta kin}$, Eq.~(\ref{eq:Exc__2}),
and neglecting all the exchange and ``standard" correlations terms.
We thus have to find a good approximation of $E_{\Delta kin}$ to proceed.
We propose to search for an approximation as a functional of the KS orbitals $\varphi^{i}_{int}$.
We adopt this approach because it provides a lot of flexibility while being fully coherent with DFT
(indeed, the KS orbitals are functionals of the internal density, i.e., $\varphi^i_{int}[\rho_{int}]$, 
as soon as they satisfy KS equations \cite{Dre90},
that can be constrained explicitly 
by use of the Optimized Effective Potential (OEP) method~\cite{Sha53a,Tal76,Kue07aR}).

\subsection{\JMcomm{The commonly used form for the c.m. correlations functional.}}
\label{sec:current_cm}

We first show how the proposed method permits us to recover the commonly used $-<\frac{\mathbf{P}^2}{2mN}>$ form for $E_{cm}$
and to understand its limitations.
We rewrite $E_{\Delta kin}$, Eq.~(\ref{eq:Exc}), in the following equivalent way
\footnote{
We mention that the square root of the delta function is not defined.
To be perfectly rigorous, we should have introduced $\lim_{\Gamma^{aux}\rightarrow\delta} \sqrt{\Gamma^{aux}(\mathbf{R})}$, 
where $\Gamma^{aux}$ is a normalized function,
instead of $\sqrt{\delta(\mathbf{R})}$.
We nevertheless use this last notation to lighten the text, 
which does not affect the conclusions.
}:
%
\begin{eqnarray}
\lefteqn{ E_{\Delta kin}[\rho_{int}] }
\nonumber\\
&&
      = \int \! d\vec{R} \delta(\mathbf{R}) (\psi_{int}|\sum_{\alpha=1}^{N-1} \frac{\tau_\alpha^2}{2\mu_\alpha}|\psi_{int}) 
      - \sum_{i=1}^{N} (\varphi^i_{int}|\frac{\vec{p}^2}{2m}|\varphi^i_{int}) 
\nonumber\\
&&
      = \int \! d\vec{R} d\vec{\xi}_1 \cdots d\vec{\xi}_{N-1} \,
      \Big(\sqrt{\delta(\mathbf{R})} \psi_{int}(\vec{\xi}_1, \cdots, \vec{\xi}_{N-1})\Big)^*
\nonumber\\
&&    \times \sum_{\alpha=1}^{N-1} \frac{\tau_\alpha^2}{2\mu_\alpha}
      \Big(\sqrt{\delta(\mathbf{R})}\psi_{int}(\vec{\xi}_1, \cdots, \vec{\xi}_{N-1})\Big)
\nonumber\\
&&  - \sum_{i=1}^{N} (\varphi^i_{int}|\frac{\vec{p}^2}{2m}|\varphi^i_{int}) 
\nonumber\\
&&
    = \int \! d\vec{r}_1 \cdots d\vec{r}_N \,
      \Big(\sqrt{\delta(\mathbf{R})} \psi_{int}(\vec{r}_1,\dots,\vec{r}_N)\Big)^*
\nonumber\\
&&    \times \Big(\sum_{i=1}^N \frac{\mathbf{p}_i^2}{2m} - \frac{\mathbf{P}^2}{2mN}\Big)
      \Big(\sqrt{\delta(\mathbf{R})}\psi_{int}(\vec{r}_1,\dots,\vec{r}_N)\Big)
\nonumber\\
&&    - \sum_{i=1}^{N} (\varphi^i_{int}|\frac{\vec{p}^2}{2m}|\varphi^i_{int}) 
.
\label{eq:Ecm_standard}
\end{eqnarray}
%
$\sqrt{\delta(\mathbf{R})}\psi_{int}(\vec{r}_1,\dots,\vec{r}_N)$
is interpreted as the c.m. frame N-body ``wavefunction"
(recall that $\psi_{int}$ has the dimension of a $(N-1)$-body wavefunction, see Eq.~(\ref{eq:psi})).
This ``wavefunction" is obviously not translationally invariant
(the $\delta(\mathbf{R})$ fixes the c.m. in position space and amounts to moving in the c.m. frame)
and antisymmetric under the exchange of two particles
(as $\psi_{int}$ is antisymmetric).
It is non null only for the $\{\vec{r}_{i}\}$ that satisfy $\mathbf{R}=\sum_{i=1}^N \vec{r}_{i} = 0$,
so that the $\{\vec{r}_{i}\}$ become the c.m. frame coordinates.

\JMcomm{
Within the internal DFT formalism, the commonly used approximation to treat the c.m. correlations
can be recovered by supposing
that the  KS Slater Determinant, denoted $\psi^{aux}$,
is a good first order approximation of the c.m. frame N-body ``wavefunction":
%
\begin{eqnarray}
\label{eq:approx_cm_wf}
\sqrt{\delta(\mathbf{R})}\psi_{int}(\vec{r}_1,\dots,\vec{r}_N)
\quad\approx\quad\
\psi^{aux}(\vec{r}_1,\dots,\vec{r}_N)
,
\end{eqnarray}
where:
$$
\psi^{aux}(\vec{r}_1,\dots,\vec{r}_N) = \frac{1}{\sqrt{N!}} \sum_{P} (-1)^p \Pi_{i=1}^N \varphi^{P(i)}_{int}(\vec{r}_i)
.
$$
($P$ are the possible permutations of the coordinates and $p$ the number of transpositions of $P$.)
%
Inserting this approximation in (\ref{eq:Ecm_standard}) and following the method described in \S\ref{sec:method_proposed}
(the ``standard" correlations are by construction neglected and the exchange terms naturally cancel),
we obtain:
%
\begin{eqnarray}
\lefteqn{ E^{}_{\Delta kin} \rightarrow E^{}_{cm}[\{\varphi^{k}_{int}\}] }
\nonumber\\
&&\quad=
- \Big(\psi^{aux}\Big|\frac{\mathbf{P}^2}{2mN}\Big|\psi^{aux}\Big)
\nonumber\\
&&\quad=
- \sum_{i=1}^N (\varphi^{i}_{int}|\frac{\mathbf{p}^2}{2mN}|\varphi^{i}_{int})
\nonumber\\
&&\quad
- \frac{1}{2mN} \sum_{i,j=1}^N
(\varphi^{i}_{int}|\mathbf{p}|\varphi^{i}_{int})
(\varphi^{j}_{int}|\mathbf{p}|\varphi^{j}_{int})
.
\label{eq:approx_cm_pot}
\end{eqnarray}
%
We recover the commonly used form for the c.m. correlations functional.
Note that in practice the term of the last line of Eq.~(\ref{eq:approx_cm_pot}) is often neglected 
to reduce the numerical cost~\cite{ben00}.
}

\JMcomm{
The internal DFT formalism permits us to shed new light on the validity of the approximation (\ref{eq:approx_cm_pot}).
It holds if and only if the approximation (\ref{eq:approx_cm_wf}) holds at least to first order.
But in general this cannot be the case because
$\psi^{aux}$ is far from being null when $\sum_{i=1}^N \vec{r}_{i} \ne 0$.
Moreover, $\psi^{aux}$ contains a c.m. vibration
typical of Slater determinants (i.e., $(\psi^{aux}|\vec{P}^n|\psi^{aux})\ne 0$ for $n\ge 2$) \cite{RS80,Schm01a},
whereas $\sqrt{\delta(\mathbf{R})}\psi_{int}$ does not contain such a vibration
(i.e., $(\psi^{int}|\vec{P}^n|\psi^{int})= 0$, $\forall n$).
Thus, we cannot expect to obtain a systematically satisfying improvement with this form~\cite{ben00}.
}

\JMcomm{
In the next section, we propose an improved form for the c.m. correlations energy functional,
where the c.m. correlations (the $\delta(\mathbf{R})$ term) appear explicitly.
}

%
%
\section{A general new form for a local c.m. correlations potential.}
\label{sec:univ}

\subsection{The idea and the result.}
\label{sec:method}
\label{sec:The functional}

We adopt a different point of view from that of \S \ref{sec:current_cm}.
We start with $E_{\Delta kin}[\rho_{int}]$ written as in Eq.~(\ref{eq:Exc__2})
(instead of Eq.~(\ref{eq:Ecm_standard}))
and do the replacement
(instead of Eq.~(\ref{eq:approx_cm_wf})):
%
\begin{eqnarray}
\label{eq:zzz2}
\psi_{int}(\vec{r}_1,\dots,\vec{r}_N) 
\quad\rightarrow\quad\
\frac{1}{\Gamma^{aux}(\vec{R})} \psi^{aux}(\vec{r}_1,\dots,\vec{r}_N)
,
\end{eqnarray}
%
%
where $\Gamma^{aux}(\vec{R})$ is any non-null one-body ``wave-function"
that implicitly depends on the number of particles $N$.
The reasons for its introduction are the following:
\begin{itemize}
\item $\psi_{int}$ has the dimension of a $(N-1)$-body wavefunction, 
whereas the KS Slater Determinant 
$\psi^{aux}(\vec{r}_1,\dots,\vec{r}_N)$
has the dimension of a $N$-body wavefunction.
Dividing $\psi^{aux}$ by $\Gamma^{aux}$ permits us to recover the correct dimension
while preserving antisymmetry.
\item The KS Slater determinant $\psi^{aux}$ contains a c.m. vibration,
whereas $\psi_{int}$ must not contain such a vibration, as already mentioned in \S\ref{sec:current_cm}.
This is not a problem from the KS point of view,
where $\psi^{aux}$ represents nothing more than an auxiliary quantity
that must only reproduce the correct $\rho_{int}$.
But if we want to replace $\psi_{int}$ by a form constructed from $\psi^{aux}$ in Eq.~(\ref{eq:Exc__2}), the c.m. vibration has to be ``subtracted" from $\psi^{aux}$.
$\Gamma^{aux}$ represents the proposed way to achieve this ``subtraction".
\item The $\delta(\vec{R})$ term, and thus the c.m. correlations, will appear explicitly
in the functional.
\item As we will see, the final result has a clear physical meaning and leads to convincing numerical results,
which shows its pertinence.
\end{itemize}
%

In the particular harmonic oscillator case (i.e., when the interaction $u$ is parabolic),
we always can achieve the separation $\psi^{aux}(\vec{r}_1,\dots,\vec{r}_{N})=\Gamma^{aux}(\vec{R})\times F(\xi_1,\dots,\xi_{N-1})$ \cite{Irv80}.
Thus, the $\Gamma^{aux}$ term introduced in Eq.~(\ref{eq:zzz2}) permits us to directly ``subtract" all the c.m. vibration contained in $\psi^{aux}$,
and $\frac{1}{\Gamma^{aux}} \psi^{aux}$ leads
to a good approximation of $\psi_{int}$.

However, in the general case, 
we do not expect $\frac{1}{\Gamma^{aux}} \psi^{aux}$ to be strictly speaking a good approximation of $\psi_{int}$.
Indeed, $\psi_{int}$ is translationally invariant whereas $\frac{1}{\Gamma^{aux}} \psi^{aux}$ is not anymore.
In other terms, $\psi^{aux}(\vec{r}_1,\dots,\vec{r}_{N})$ cannot be separated into $\Gamma^{aux}(\vec{R})\times F(\xi_1,\dots,\xi_{N-1})$.
This is not a problem because in every integral where $\psi_{int}$ appears (that represent observables),
a $\delta(\vec{R})$ term that breaks translational invariance also appears explicitly.
What we expect is that the replacement (\ref{eq:zzz2}), i.e., the introduction of $\Gamma^{aux}$,
allows sufficient flexibility to lead to a satisfying result for both $E_{cm}$ and $\rho_{int}$.
Then, even if the ``subtraction" is not ``direct" because there is no separation of the c.m. motion,
it is ``indirect" because it leads to the correct final result.
Note that because of the $\delta(\vec{R})$ that appears in all integrals that represent observables
only the values and variations of $\Gamma^{aux}$ around $\vec{R}=\vec{0}$ can contribute.


We now insert the approximation (\ref{eq:zzz2}) in (\ref{eq:Exc__2}) and 
keep only the real part of the result
(indeed, the straightforward result
leads to a complex $E^{}_{\Delta kin}$ in the general case,
that is fundamentally due to the fact that
the form (\ref{eq:zzz2}) cannot be rewritten as
a function of the $\{\xi_\alpha\}$ only).
We then obtain an approximation of the exact $E^{}_{\Delta kin}$
where the ``standard" correlations have been neglected by construction.
As discussed in \S\ref{sec:method_proposed}, it remains to neglect the exchange terms to obtain $E_{cm}$.
The calculation is detailed in Appendix \ref{app:app1}.
The final result is:
\begin{widetext}
\begin{eqnarray}
&&E^{}_{\Delta kin} \rightarrow E^{}_{cm}[\{\varphi^{k}_{int}\}] =
\nonumber\\
&&\quad
-\frac{\hbar^2}{2m}
\sum_{i=1}^N \int d\vec{r} \hspace{1mm}
\varphi^{i*}_{int}(\vec{r})\Delta_{\vec{r}}\varphi^{i}_{int}(\vec{r})
\times
\Big( \frac{1}{|\Gamma^{aux}(\vec{0})|^2} \int d\vec{r}' \hspace{1mm}
|\varphi^{l\ne i}_{int}(\vec{r}')|^2 
\times
f_{i,l\ne i}[\{\varphi^{k\ne i,l}_{int}\}](\vec{r}+\vec{r}') - 1 \Big)
\nonumber\\
&&\quad
-\frac{\hbar^2}{2mN} 
\hspace{1mm}
\frac{1}{\Gamma^{aux*}(\vec{0})} \Delta_\vec{R} \frac{1}{\Gamma^{aux}(\vec{R})}\Big|_{\vec{R}=\vec{0}} 
\times
\int d\vec{r} \hspace{1mm} 
|\varphi^{i}_{int}(\vec{r})|^2
\int d\vec{r}' \hspace{1mm}
|\varphi^{l\ne i}_{int}(\vec{r}')|^2 
\times f_{i,l\ne i}[\{\varphi^{k\ne i,l}_{int}\}](\vec{r}+\vec{r}')
\nonumber\\
&&\quad
\JMcomm{
+ \hspace{1mm}
PureImaginary[\{\varphi^{k}_{int}\}]
}
,
\label{eq:E12_4-}
\end{eqnarray}
\end{widetext}
where the functional $PureImaginary$
%
%
counteracts the pure imaginary part of
the second and third lines of Eq.~(\ref{eq:E12_4-}) and becomes null in the real (stationary) case.
We keep this functional for the general (time-dependent) case.

The ``two-particle c.m. correlations functional" is defined as:
\begin{eqnarray}
\lefteqn{ f_{i,l\ne i}[\{\varphi^{k\ne i,l}_{int}\}](\tilde{\vec{r}}) = }
\label{eq:f}\\
&& N^D \int \Pi_{\stackrel{j=1}{j\ne i,l}}^N d\vec{r}_j \delta\big(\sum_{\stackrel{k=1}{k\ne i,l}}^N \vec{r}_k + \tilde{\vec{r}}\big) 
\Pi_{\stackrel{j=1}{j\ne i,l}}^N 
|\varphi^{j}_{int}(\vec{r}_j)|^2
,
\nonumber
\end{eqnarray}
where D is the dimension in which the calculation is done (D=1, 2 or 3).
In the following, we note $f_{i,l\ne i}$ instead of $f_{i,l\ne i}[\{\varphi^{k\ne i,l}_{int}\}]$
to lighten the notations.
The meaning and properties of this functional will be detailed in \S \ref{sec:c.m. cor functional}.

The potentials $U_{cm}^{l}$ corresponding to $E_{cm}$ are defined by ($l=1...N$)
\begin{widetext}
\begin{eqnarray}
U_{cm}^{l}(\vec{r}) \varphi^{l}_{int}(\vec{r}) =
\frac{\delta E^{}_{cm}[\{\varphi^{k}_{int}\}]}{\delta \varphi^{l*}_{int}(\vec{r})}
&=&
-\frac{\hbar^2}{2m}
\Big\{
\Delta_{\vec{r}}\varphi^{l}_{int}(\vec{r})
\times \Big( \frac{1}{|\Gamma^{aux}(\vec{0})|^2} \int d\vec{r}'
|\varphi^{m\ne l}_{int}(\vec{r}')|^2 
f_{l,m\ne l}
(\vec{r}+\vec{r}') -1 \Big)
\nonumber\\
&&
\quad\quad\quad
+
\frac{1}{|\Gamma^{aux}(\vec{0})|^2} \varphi^{l}_{int}(\vec{r}) \sum_{\stackrel{i=1}{i\ne l}}^N \int d\vec{r}' \hspace{1mm}
\varphi^{i*}_{int}(\vec{r}') \Delta_{\vec{r}'}\varphi^{i}_{int}(\vec{r}')
\times
f_{i,l\ne i}
(\vec{r}+\vec{r}')
\Big\}
\nonumber\\
&&
-\frac{\hbar^2}{2mN} 
\hspace{1mm}
\frac{1}{\Gamma^{aux*}(\vec{0})} \Delta_\vec{R} \frac{1}{\Gamma^{aux}(\vec{R})}\Big|_{\vec{R}=\vec{0}}
\times
\varphi^{l}_{int}(\vec{r})
\int d\vec{r}' 
|\varphi^{m\ne l}_{int}(\vec{r}')|^2 
\times f_{l,m\ne l}
(\vec{r}+\vec{r}')
\nonumber\\
&&
\JMcomm{
+ \hspace{1mm}
\frac{\delta}{\delta \varphi^{l*}_{int}(\vec{r})} PureImaginary[\{\varphi^{k}_{int}\}]
}
,
\label{eq:cm_pot_2}
\end{eqnarray}
\end{widetext}
where the last line is obviously null in the real (stationary) case.
%
Note that the potentials $U_{cm}^{l}$ are not the same for all states. This is due to the fact that $E_{cm}[\{\varphi^{k}_{int}\}]$
is orbital-dependent
which requires extra measures to recover a common potential or,
equivalently, to preserve orthonormalization.
A way to overcome this problem and remain fully coherent with DFT is to use the OEP method
which permits us to find the potential common to all states
that reproduces most accurately the effect of the $U_{cm}^{l}$ potentials.
We refer the reader to Refs.~\cite{Sha53a,Tal76,Kue07aR} for the exhaustive equations.
As the full OEP result is very costly numerically, it is often simplified. 
The Krieger-Li-Iafrate (KLI) approach is a popular
approach and, in a further step of simplification,
the Slater approximation \cite{OEP1,OEP2} is used.
As our goal is to find a numerically inexpensive form for the local c.m. potential, we detail hereafter
only the Slater approximation:
\begin{eqnarray}
\label{eq:cm_pot2}
U^{Slat}_{cm}(\vec{r}) = 
\frac{1}{\rho_{int}(\vec{r})} \sum_{i=l}^{N} |\varphi^{l}_{int}(\vec{r})|^2 U_{cm}^{l}(\vec{r})
.
\end{eqnarray}
%
%

\subsection{Properties of $f_{i,l\ne i}$ and numerical considerations.}
\label{sec:c.m. cor functional}

The definition (\ref{eq:f}) of the two-particle c.m. correlations functional $f_{i,l\ne i}$ shows that:
\begin{itemize}
\item it is real and has the dimension of a density,
\item it is normalized to $N^D$, i.e. $\int d\tilde{\vec{r}} f_{i,l\ne i}(\tilde{\vec{r}}) = N^D$,
\item $\lim_{\tilde{\vec{r}}\rightarrow \pm\infty} f_{i,l\ne i}(\tilde{\vec{r}})=0$,
\item it is a ``multiconvolution" of all single densities, unless these are associated to orbitals $i$ and $l$.
\end{itemize}
The first three points permit us to make explicit the physical meaning of $\frac{1}{N^D} f_{i,l\ne i}(\vec{r}+\vec{r}')$:
it is the probability that particle $l\ne i$ has position $\vec{r}'$, given that particle $i$ has position $\vec{r}$.
Indeed, because of the c.m. correlations, the positions of those particles are not independent;
every single orbital $\varphi^{i}_{int}$ is coupled to every single orbital $\varphi^{l\ne i}_{int}$ through $f_{i,l\ne i}$.
This coupling appears in the c.m. correlations energy (\ref{eq:E12_4-}) and potentials (\ref{eq:cm_pot_2}).

To better understand this coupling, note that $f_{i,l\ne i}$ can be rewritten as:
\begin{eqnarray}
\lefteqn{ f_{i,l\ne i}
(\tilde{\vec{r}}) =}
\nonumber\\
&& 2^D \delta(\tilde{\vec{r}}) \quad if \quad N = 2,
\nonumber\\
&&
3^D |\varphi^{k\ne i, l}_{int}\big(- \tilde{\vec{r}}\big)|^2
\quad if \quad N = 3,
\nonumber\\
&& N^D \int \Pi_{\stackrel{j=1}{j\ne i,l,m}}^N d\vec{r}_j \Pi_{\stackrel{j=1}{j\ne i,l,m}}^N 
|\varphi^{j}_{int}(\vec{r}_j)|^2
\nonumber\\
&&\hspace{2cm} \times  |\varphi^{m}_{int}\big(-\sum_{\stackrel{k=1}{k\ne i,l,m}}^N \vec{r}_k - \tilde{\vec{r}}\big)|^2
\quad if \quad N \ge 4,
\label{eq:fN>4}
\\
&&
\dots
\nonumber\\
&& Constant,
\quad for \hspace{1.5mm} very \hspace{1.5mm} large \hspace{1.5mm} N \hspace{1.5mm} (limit\hspace{1.5mm} of\hspace{1.5mm} a\hspace{1.5mm} Fermi\hspace{1.5mm} gas).
\nonumber
\end{eqnarray}
We see that, in the two-particle case, $f_{1,2}$ is proportional to the steep delta function. Thus, if particle $1$
has position $\vec{r}$, particle $2$ will have position -$\vec{r}$, so that the c.m. remains stuck at $\vec{R}=\vec{0}$.
In the three-particle case, $f_{i,l\ne i}$ has a larger width, because the introduction of a third particle
allows more freedom to the motion of the two other particles, while preserving $\vec{R}=\vec{0}$.
For $N \ge 4$, the width of $f_{i,l\ne i}$ will increase as $N$ grows,
because of the multiconvolution form of $f_{i,l\ne i}$.
Indeed, a larger number of particles allows more liberty to the motion of two of them
while preserving $\vec{R}=\vec{0}$.
For very large $N$, the system tends to a Fermi gas, so that $f_{i,l\ne i}$
tends to become constant and delocalized in the whole space, i.e., the motions of the particles tend to become independent.
The c.m. correlations can then be neglected, as expected.

Practically speaking, we see that the numerical cost of the whole scheme lies in the calculation of $f_{i,l\ne i}$ for $N\ge 4$, 
i.e., the calculation of the multiconvolution of Eq.~(\ref{eq:fN>4}).
At first sight, it seems to be disadvantageous for large $N$.
But a mathematical property of the convolutions under Fourier transforms makes it manageable.
In Appendix \ref{app:app2}, we recall the so-called ``multiconvolution theorem".
Its direct application to $f_{i,l\ne i}$ for $N\ge 4$ gives:
$$
f_{i,l\ne i}
(\tilde{\vec{r}}) =
N^D \times \mathcal{T}^{-1} \Big[ \Pi_{\stackrel{k=1}{k\ne i,l}}^N \mathcal{T}[|\varphi^{k}_{int}|^2] \Big] (-\tilde{\vec{r}})
,
$$
where $\mathcal{T}$ denotes the Fourier transform as defined in Appendix \ref{app:app2}, Eq.~(\ref{eq:FT}).
This permits us to drastically shorten the numerical calculation of $f_{i,l\ne i}$
which becomes manageable even for large systems.
Indeed, once all the $\mathcal{T}[|\varphi^{k}_{int}|^2]$ are calculated, $f_{i,l\ne i}$ is given by
the inverse Fourier transform of their direct product,
so that the numerical cost of $f_{i,l\ne i}$ equals the numerical cost of $(N+1)$ Fast Fourier Transforms when $N \ge 4$.
%



\subsection{Properties of $\Gamma^{aux}$ and numerical considerations.}
\label{sec:Gamma_aux}

To completely characterize $E^{}_{cm}$, we still need to characterize
the values of $|\Gamma^{aux}(\vec{0})|^2$ and $\frac{1}{\Gamma^{aux*}(\vec{0})} \Delta_\vec{R} \frac{1}{\Gamma^{aux}(\vec{R})}\Big|_{\vec{R}=\vec{0}}$,
see Eq.~(\ref{eq:E12_4-}).


\subsubsection{First step: value of $|\Gamma^{aux}(\vec{0})|^2$.}
\label{sec:1st_step}
$|\Gamma^{aux}(\vec{0})|^2$ is imposed by the normalization condition on the approximation (\ref{eq:zzz2}) we used for $\psi_{int}$
\footnote{
The last line of Eq.~(\ref{eq:normal}) is obtained introducing the form (\ref{eq:zzz2}) for $\psi_{int}$ in the first line of Eq.~(\ref{eq:normal})
and neglecting the exchange terms.
}:
\begin{eqnarray}
&&1=(\psi_{int}|\psi_{int})=\int \! d\vec{r}_1 \cdots d\vec{r}_N \; \delta(\mathbf{R}) |\psi_{int}(\vec{r}_1, \ldots , \vec{r}_N)|^2 
\nonumber\\
&&\quad\quad \Rightarrow \quad 
\label{eq:normal}
\\
&& |\Gamma^{aux}(\vec{0})|^2 =  
\int d\vec{r} \hspace{1mm} d\vec{r}' \hspace{1mm}
|\varphi^{i}_{int}(\vec{r})|^2
|\varphi^{l\ne i}_{int}(\vec{r}')|^2 
f_{i,l\ne i}
(\vec{r}+\vec{r}')
.
\nonumber
\end{eqnarray}
Numerically speaking, this condition will be satisfied self-consistently, starting from a reasonable initial value for $|\Gamma^{aux}(\vec{0})|^2$
and rescaling it at every numerical loop so that it satisfies the last line of Eq.~(\ref{eq:normal}).

\subsubsection{Second step: value of $\frac{1}{\Gamma^{aux*}(\vec{0})} \Delta_\vec{R} \frac{1}{\Gamma^{aux}(\vec{R})}\Big|_{\vec{R}=\vec{0}}$.}
\label{sec:2nd_step}

To characterize this value,
we define a pertinent continuous set of normalized functions $\{\Gamma^{aux}(\vec{R})\}$ that are twice derivable.
We then choose at each numerical step the particular $\Gamma^{aux}$ function of the set
whose norm squared in $\vec{R}=\vec{0}$ is the one that has been obtained in the first step (the set should unambiguously define this value).
Then we calculate $\frac{1}{\Gamma^{aux*}(\vec{0})} \Delta_\vec{R} \frac{1}{\Gamma^{aux}(\vec{R})}\Big|_{\vec{R}=\vec{0}}$ with it.
This permits us to completely define the c.m. correlation energy (\ref{eq:E12_4-}) and potential (\ref{eq:cm_pot_2})
without introducing any free parameter.

We mention a particular relation that should satisfy the $\Gamma^{aux}$ functions choosen to constitute the set.
Recall that $\Gamma^{aux}$ depends implicitly on $N$.
As demonstrated in Appendix \ref{app:app3}, $\Gamma^{aux}$
should satisfy the following properties as $N$ increases:
\begin{eqnarray}
&&\lim_{N\rightarrow +\infty} |\Gamma^{aux}(\vec{0})|^2 \rightarrow  +\infty
,
\label{eq:lim_gamma_aux}
\\
&&\lim_{N\rightarrow +\infty}
\frac{1}{N}\times\frac{1}{\Gamma^{aux*}(\vec{0})} \Delta_\vec{R} \frac{1}{\Gamma^{aux}(\vec{R})}\Big|_{\vec{R}=\vec{0}}\times |\Gamma^{aux}(\vec{0})|^2
\rightarrow 0
.
\nonumber
\end{eqnarray}
The remaining task is to choose a pertinent continuous set of $\{\Gamma^{aux}(\vec{R})\}$ that satisfies those properties.

\subsubsection{Practical proposition.}

The most simple \cite{Pei62} set for $\Gamma^{aux}$
that meets all the previously mentioned criteria
(and is exact in the case where the interaction $u$ is parabolic)
is the Gaussian set:
\begin{eqnarray}
\label{eq:gamma_aux_3}
\Gamma^{aux}(\vec{R}) =
\Big( \frac{K_N}{\pi} \Big)^{D/4} \exp\Big\{-\frac{K_N}{2}\sum_{i=1}^D R_i^2\Big\}
,
\end{eqnarray}
where $R_i$ are the coordinates of $\mathbf{R}$ in $D$ dimensions and $K_N$ is the parameter that defines $\Gamma^{aux}$ for every given $N$.
With this form:
\begin{eqnarray}
&&|\Gamma^{aux}(\vec{0})|^2=\Big( \frac{K_N}{\pi} \Big)^{D/2},
\label{eq:gamma_aux_10}\\
&&\frac{1}{\Gamma^{aux*}(\vec{0})} \Delta_\vec{R} \frac{1}{\Gamma^{aux}(\vec{R})}\Big|_{\vec{R}=\vec{0}}
=
\pi^{D/2} \times D \times K_N^{1-D/2}
\label{eq:gamma_aux_11}.
\end{eqnarray}
For a given system composed of $N$ particles, $|\Gamma^{aux}(\vec{0})|^2$ is still obtained at each numerical loop with the first step (\S\ref{sec:1st_step}),
that defines $K_N$ by Eq.~(\ref{eq:gamma_aux_10}) and
$\frac{1}{\Gamma^{aux*}(\vec{0})} \Delta_\vec{R} \frac{1}{\Gamma^{aux}(\vec{R})}\Big|_{\vec{R}=\vec{0}}$ 
by Eq.~(\ref{eq:gamma_aux_11}).

We underline that the choice of a Gaussian set for $\Gamma^{aux}$
absolutely does not constrain the $\varphi^i_{int}$ to show a Gaussian behavior (even asymptotically).
Indeed, it simply gives a method to define the value of $\frac{1}{\Gamma^{aux*}(\vec{0})} \Delta_\vec{R} \frac{1}{\Gamma^{aux}(\vec{R})}\Big|_{\vec{R}=\vec{0}}$
given the value of $|\Gamma^{aux}(\vec{0})|^2$,
where only the behaviour of $\Gamma^{aux}$ around $\vec{R}=\vec{0}$ enters into account.
The numerical results presented thereafter will show that the Gaussian set choice gives good results. 
Nevertheless, the search for other sets, i.e., with other variations around $\vec{R}=\vec{0}$,
should be continued to obtain the most precise description of self-bound systems in fully realistic calculations.
This investigation goes beyond the scope of this paper.

\subsubsection{Initial condition.}

With this method, there is no need to analytically define $K_N$ as a function of $N$;
$K_N$ is obtained numerically for every given $N$ as indicated previously.
It would nevertheless be interesting to obtain an approximate analytical form
to start the numerical iterations with a pertinent initial condition.
To that aim, 
we note that the conditions (\ref{eq:lim_gamma_aux}),
together with the equalities (\ref{eq:gamma_aux_10}) and (\ref{eq:gamma_aux_11})
imply the following conditions on $K_N$:
\begin{eqnarray}
&&
\lim_{N\rightarrow +\infty} K_N^{D/2}  \rightarrow +\infty,
\nonumber\\
&&\lim_{N\rightarrow +\infty} \frac{1}{N} K_N \rightarrow 0
.
\end{eqnarray}
A straightforward form for $K_N$ that satisfies those two constraints is:
\begin{eqnarray}
\label{eq:param_K}
K_N = A\times N^{a} \quad , \quad where \hspace{1.5mm} 0<a<1
.
\end{eqnarray}
In practice, $a\in[0.6;0.9]$ should be reasonable choice in nuclear physics~\footnote{
The corresponding energy associated to $\Gamma^{aux}$ is
$
E_{\Gamma^{aux}} = -\frac{\hbar^2}{2Nm}(\Gamma^{aux}|\Delta_\vec{R}|\Gamma^{aux}) =
\frac{\hbar^2}{2Nm} \frac{D}{4} K_N =
\frac{\hbar^2}{2m} \frac{D}{4} A\times N^{a-1}
$.
$E_{\Gamma^{aux}}$ is proportional to $N^{a-1}$.
Even if $E_{\Gamma^{aux}}$ has, strictly speaking, no physical meaning, is is reasonable to assume that its variation
according to $N$ should approximately be proportional to
the variation of the energy associated to the c.m. vibration
obtained in mean-field-like calculation, see Ref.~\cite{ben00}.
For the nuclear case, the c.m. vibration energy evaluated for harmonic oscillator states 
is proportional to $N^{-1/3}$; this variation can be reproduced with $a = 2/3$.
The c.m. correlation energy evaluated with a \textit{a posteriori} fit with mean-field-like calculations 
is proportional to $N^{-0.2}$; this variation can be reproduced with $a\approx 0.8$~\cite{ben00}.
}.

\subsection{``By-products".}

\subsubsection{An explicit density functional for Fermions.}
\label{sec:c.m. cor functional rho}

The functional proposed in \S\ref{sec:The functional} 
is by construction not an explicit functional of $\rho_{int}$ (but an orbital dependent functional).
It is well-suited for stationary calculations
but not for time-dependent ones because
the Slater approximation
does not permit us to preserve energy conservation 
as this approximation is not perfectly variational~\cite{Mes11-2}.
Only the full time-dependent OEP result \cite{Ull95a} will achieve energy conservation
but at the price of a much larger numerical cost.
It would be interesting to find an explicit functional of $\rho_{int}$ that would overcome this time-dependent case problem.

In this section, we propose a further step of approximation that will allow us to obtain such a functional.
%
We do the Local Density Approximation (LDA) on the result of \S \ref{sec:The functional},
which consists of assuming that the system is {locally} homogeneous~\cite{Dre90,Par89}.
Despite its simplicity, this approximation has proven to be very satisfying to describe a wide range of systems,
and not only large ones~\cite{RS80,Dre90}.
%
To make the LDA, we first make the replacement:
\begin{eqnarray}
|\varphi^{i}_{int}(\vec{r})|^2 
&\rightarrow& 
\frac{1}{N} \rho_{int}(\vec{r})
\end{eqnarray}
everywhere the single density terms $|\varphi^{i}_{int}|^2$ appear into $E_{cm}$, Eq.~(\ref{eq:E12_4-}).
In the obtained functional, the only remaining term that is not an explicit functional of $\rho_{int}$ is
$\sum_{i=1}^N \varphi^{i*}_{int}(\vec{r})\Delta_{\vec{r}}\varphi^{i}_{int}(\vec{r})$.
As we consider a system composed of Fermions, we can make the Thomas-Fermi approximation \cite{Dre90}, i.e., the replacement:
\begin{eqnarray}
\sum_{i=1}^N \varphi^{i*}_{int}(\vec{r})\Delta_{\vec{r}}\varphi^{i}_{int}(\vec{r})
&\rightarrow&
- \frac{3}{5}C \rho_{int}^{5/3}(\mathbf{r})
,
\end{eqnarray}
where
$C=(\frac{3\pi^2}{\gamma})^{2/3}$
and $\gamma$ is the degeneracy.
%
%
We obtain as a final result the c.m. correlations energy written as an explicit functional of $\rho_{int}$:
%
\begin{eqnarray}
\label{eq:E12_4---LDA}
\lefteqn{E^{LDA}_{cm}[\rho_{int}] =}
\\
&&\frac{\hbar^2}{2m}
\int d\vec{r} \hspace{1mm}
\frac{3}{5} C \rho_{int}^{5/3}(\mathbf{r})
\nonumber\\
&&\quad\times \Big( \frac{1}{|\Gamma^{aux}(\vec{0})|^2} \int d\vec{r}' \hspace{1mm}
\frac{1}{N} \rho_{int}(\vec{r}')
\times
f_2[\rho_{int}](\vec{r}+\vec{r}') - 1 \Big)
\nonumber\\
&&\quad
-\frac{\hbar^2}{2mN} 
\hspace{1mm}\Re e \Big(
\frac{1}{\Gamma^{aux*}(\vec{0})} \Delta_\vec{R} \frac{1}{\Gamma^{aux}(\vec{R})}\Big|_{\vec{R}=\vec{0}}
\Big)
\nonumber\\
&&\quad\times 
\int d\vec{r} \hspace{1mm} 
\frac{1}{N} \rho_{int}(\vec{r})
\int d\vec{r}' \hspace{1mm}
\frac{1}{N} \rho_{int}(\vec{r}')
\times f_2[\rho_{int}](\vec{r}+\vec{r}')
\nonumber
,
\end{eqnarray}
%
where the ``two-particle {average} c.m. correlations functional" is defined by:
\begin{eqnarray}
\lefteqn{ f_2
(\tilde{\vec{r}}) = }
\nonumber\\
&& 2^D \delta(\tilde{\vec{r}}) \quad if \quad N = 2,
\nonumber\\
&&
3^D \frac{1}{N} \rho_{int}\big(- \tilde{\vec{r}} \big)
\quad if \quad N = 3,
\nonumber\\
&& N^D \frac{1}{N^{N-2}} \int 
d\vec{r}_1 \dots d\vec{r}_{N-3} \hspace{1mm} \rho_{int}(\vec{r}_1) \times \dots \times 
\nonumber\\
&&\quad\quad\quad \rho_{int}(\vec{r}_{N-3}) \hspace{1mm}
\rho_{int}\big(-\sum_{k=1}^{N-3} \vec{r}_k - \tilde{\vec{r}} \big)
\quad if \quad N \ge 4,
\label{eq:f_LDA_2}\\
&&
\dots
\nonumber\\
&& Constant,
\quad for \hspace{1.5mm} very \hspace{1.5mm} large \hspace{1.5mm} N \hspace{1.5mm} (limit\hspace{1.5mm} of\hspace{1.5mm} a\hspace{1.5mm} Fermi\hspace{1.5mm} gas).
\nonumber
\end{eqnarray}
(We note $f_2$ instead of $f_2[\rho_{int}]$ to lighten the notation.)
Similar considerations to those of \S \ref{sec:c.m. cor functional}
permit us to interpret $\frac{1}{N^D} f_2(\vec{r}+\vec{r}')$ as
the average probability that one particle has position $\vec{r}'$ given that another particle has position $\vec{r}$.
%

The corresponding unique c.m. correlations potential is
given by:
%
%
\begin{widetext}
\begin{eqnarray}
\frac{\delta E^{LDA}_{cm}[\rho_{int}]}{\delta \rho_{int}(\vec{r})} = U^{LDA}_{cm}[\rho_{int}](\vec{r}) &=& 
\frac{\hbar^2}{2m}
\Big\{
C \rho_{int}^{2/3}(\mathbf{r})
\times \Big( \frac{1}{|\Gamma^{aux}(\vec{0})|^2} \int d\vec{r}'
\frac{1}{N} \rho_{int}(\vec{r}')
f_2
(\vec{r}+\vec{r}') -1 \Big)
\label{eq:cm_pot_LDA}\\
&&
\quad\quad\quad
+
\frac{1}{|\Gamma^{aux}(\vec{0})|^2}
\int d\vec{r}' \hspace{1mm}
\frac{3}{5} C \rho_{int}^{5/3}(\mathbf{r}') \times \frac{N-1}{N}
f_2
(\vec{r}+\vec{r}')
\Big\}
\nonumber\\
&&
-\frac{\hbar^2}{2mN} 
\hspace{1mm}\JMcomm{\Re e \Big(}
\frac{1}{\Gamma^{aux*}(\vec{0})} \Delta_\vec{R} \frac{1}{\Gamma^{aux}(\vec{R})}\Big|_{\vec{R}=\vec{0}}
\JMcomm{\Big)\times}
\int d\vec{r}' 
\frac{1}{N} \rho_{int}(\vec{r}')
\times f_2
(\vec{r}+\vec{r}')
.
\nonumber
\end{eqnarray}
\end{widetext}
%

Still, we see that the numerical cost lies in the calculation of $f_2$ for $N\ge 4$.
To reduce this cost, we use the ``multiconvolution theorem" recalled in Appendix \ref{app:app2}.
Its direct application to $f_2$ for $N\ge 4$ gives
(using the definition Eq.~(\ref{eq:FT}) for the Fourier transform $\mathcal{T}$):
$$
f_2(\tilde{\vec{r}}) =
N^D \frac{1}{N^{N-2}} \times \mathcal{T}^{-1} \Big[ ( \mathcal{T}[\rho_{int}] )^{N-2} \Big] (-\tilde{\vec{r}})
.
$$
This permits us to speed up drastically the numerical calculation of $f_2$
which becomes manageable even for very large systems.
Indeed, we simply calculate $\mathcal{T}[\rho_{int}]$, raise it to power ($N-2$) and calculate
its inverse Fourier transform. Thus, the numerical cost of the calculation of $f_2$ 
is equal to two Fast Fourier Transforms for all $N \ge 4$.

Moreover, this scheme is perfectly variational, contrary to that of \S \ref{sec:The functional}, 
and thus is suitable for stationary calculations as well as for time-dependent ones
(it will achieve energy conservation if time-independent $v_{int}$ is used).

\subsubsection{An explicit density functional for Bosons.}
\label{sec:bosons}

Until now, we have only considered systems of Fermions.
The c.m. correlations energy functional for Bosons condensates
is obtained by replacing $\varphi^{i}_{int} \rightarrow \varphi_{int}$, thus $\rho_{int}=N|\varphi_{int}|^2$, in (\ref{eq:E12_4-}).
Setting $\varphi_{int}=\sqrt{\rho_{int}/N}$,
we obtain:
\begin{widetext}
\begin{eqnarray}
\label{eq:E12_4---bosons}
E_{cm}[\rho_{int}] &=&
-\frac{\hbar^2}{2m}
\int d\vec{r} \hspace{1mm}
\JMcomm{
\sqrt{\rho_{int}(\mathbf{r})}\Delta_{\vec{r}}\sqrt{\rho_{int}(\mathbf{r})}\times
}
\hspace{1mm} \Big( \frac{1}{|\Gamma^{aux}(\vec{0})|^2} \int d\vec{r}' \hspace{1mm}
\frac{1}{N} \rho_{int}(\vec{r}')
\times
f_2
(\vec{r}+\vec{r}') - 1 \Big)
\\
&&
-\frac{\hbar^2}{2mN} 
\hspace{1mm}\JMcomm{\Re e \Big(}
\frac{1}{\Gamma^{aux*}(\vec{0})} \Delta_\vec{R} \frac{1}{\Gamma^{aux}(\vec{R})}\Big|_{\vec{R}=\vec{0}}
\JMcomm{\Big)\times}
\int d\vec{r} \hspace{1mm} 
\frac{1}{N} \rho_{int}(\vec{r})
\int d\vec{r}' \hspace{1mm}
\frac{1}{N} \rho_{int}(\vec{r}')
\times f_2
(\vec{r}+\vec{r}')
\nonumber
,
\end{eqnarray}
%
where $f_2$ is defined as in Eq.~(\ref{eq:f_LDA_2})
and the corresponding c.m. correlations potential is given by
%
\begin{eqnarray}
\frac{\delta E_{cm}[\rho_{int}]}{\delta \rho_{int}(\vec{r})} = U_{cm}[\rho_{int}](\vec{r}) &=& 
-\frac{\hbar^2}{2m}
\Big\{
\JMcomm{
\frac{1}{\sqrt{\rho_{int}(\mathbf{r})}}\Delta_{\vec{r}}\sqrt{\rho_{int}(\mathbf{r})}\times
}
\Big( \frac{1}{|\Gamma^{aux}(\vec{0})|^2} \int d\vec{r}'
\frac{1}{N} \rho_{int}(\vec{r}')
f_2
(\vec{r}+\vec{r}') -1 \Big)
\nonumber\\
&&
\quad\quad\quad
+
\frac{1}{|\Gamma^{aux}(\vec{0})|^2}
\int d\vec{r}' \hspace{1mm}
\sqrt{\rho_{int}(\vec{r})}\Delta_{\vec{r}}\sqrt{\rho_{int}(\vec{r})} \times \frac{N-1}{N}
f_2
(\vec{r}+\vec{r}')
\Big\}
\nonumber\\
&&
-\frac{\hbar^2}{2mN} 
\hspace{1mm}\JMcomm{\Re e \Big(}
\frac{1}{\Gamma^{aux*}(\vec{0})} \Delta_\vec{R} \frac{1}{\Gamma^{aux}(\vec{R})}\Big|_{\vec{R}=\vec{0}}
\JMcomm{\Big)\times}
\int d\vec{r}' 
\frac{1}{N} \rho_{int}(\vec{r}')
\times f_2
(\vec{r}+\vec{r}')
.
\label{eq:cm_pot_bosons}
\end{eqnarray}
\end{widetext}
\JMcomm{
This potential is common to all states,
is an explicit functional of $\rho_{int}$,
and is strictly variational (so that it may be used in the time-dependent case).
}

%
%
\section{Numerical results.}

We consider 1D calculations
which will allow us to
better understand some features of the internal DFT formalism and more easily include various particle-particle interactions.
%

\subsection{Model system composed of two different particles with a strong interaction.}
\label{app:model0}

\subsubsection{The model and the benchmark.}
\label{app:model}

We consider a self-bound system composed of two different particles,
to maximize the c.m. correlations effects.
%
We suppose that the two particles have the same mass $m$ and are coupled by a strong interaction, which models features of a proton and a neutron.
The first particle has laboratory coordinates $r^{(1)}$, $p^{(1)}$, and the second particle has laboratory coordinate $r^{(2)}$, $p^{(2)}$.
The reduced mass is $\mu=m/2$, and the Jacobi coordinates are $\xi=r^{(1)}-r^{(2)}$, $\tau=p^{(1)}-p^{(2)}$.
We suppose that the interaction between the two particles is parabolic (harmonic oscillator)
so that the laboratory Hamiltonian is
$
H=\sum_{i=1}^2\frac{{p^{(i)}}^2}{2m} +\frac{1}{4}m\omega^2(r^{(1)}-r^{(2)})^2
$,
and the internal Hamiltonian is:
\begin{eqnarray}
H_{int}=\frac{\tau^2}{2\mu}+\frac{1}{2}\mu\omega^2\xi^2
\label{eq:H_ho}
.
\end{eqnarray}
Its ground state can be written analytically
($\psi_{int}$ should not be anti-symmetrized because we deal with two different particles):
\begin{eqnarray}
\label{eq:psi_int}
\psi_{int}(\xi)=\Big( \frac{\mu\omega}{\pi\hbar} \Big)^{\frac{1}{4}} \exp\Big\{-\frac{1}{2}\frac{\mu\omega}{\hbar}\xi^2\Big\}
.
\end{eqnarray}
The corresponding energy is $E_{int}=\frac{1}{2}\hbar\omega$ and the c.m. frame one-body densities for
each kind of particle are ($R=(r^{(1)}+r^{(2)})/2$; $l=1..2$)~\cite{Mes11}:
\begin{eqnarray}
\label{eq:rho_int}
\rho^{(l)}_{int}(r)
   &=&  \int \! dr^{(1)}dr^{(2)} \delta(R)
      |\psi_{int}(r^{(1)}-r^{(2)})|^2
      \delta \big( r - (r^{(l)}-R) \big)\,
\nonumber\\
&=& 2 |\psi_{int}(2 r)|^2 = \sqrt{ \frac{2m\omega}{\pi\hbar} } \exp\Big\{-\frac{2m\omega}{\hbar}r^2\Big\}
.
\end{eqnarray}
This is our benchmark.

It can be shown analytically using a harmonic oscillator basis that the Hartree (H) solution
(there is no exchange because the two particles are different) 
leads to $E_{int}=\frac{1}{\sqrt{2}}\hbar\omega$ and
$\rho^{(l)}_{int}(r)= \sqrt{ \frac{m\omega}{\sqrt{2}\pi\hbar} } \exp(-\frac{m\omega}{\sqrt{2}\hbar}r^2)$.
Thus the H energy is $2/\sqrt{2}$ ($\approx 1.4$) times more important than that of the benchmark,
and the density is 1.7 times more spread.
The H solution is much more delocalized than the benchmark
because the c.m. correlations are neglected
\footnote{
The c.m. correlations tend to 
localize the densities compared to the independent particle approximation
which can be understood as follows:
if c.m. correlations are taken into account, when one particle moves in a direction where the potential well is higher,
the other one will have to move in the opposite direction where the potential well is also higher.
The first particle will thus feel the repulsion present in the independent particle approximation, 
but also the repulsion felt by the second particle through the c.m. correlations.
}.

\subsubsection{The internal DFT exact functional.}
\label{app:exactDFT}

Applying the multicomponent internal DFT formalism developed in Ref.~\cite{Mes11}
(whose equations have a relatively similar form than the ``one kind of particle" internal DFT ones recalled in \S \ref{sec:intDFTrecall}),
we can rewrite the internal energy
($\varphi^{(1)}_{int}$ and $\varphi^{(2)}_{int}$ being the KS orbitals):
\begin{eqnarray}
E_{int}[{\rho}^{(1)}_{int},{\rho}^{(2)}_{int}] &=& 
\sum_{l=1}^2 (\varphi^{(l)}_{int}|\frac{{p}^2}{2m}|\varphi^{(l)}_{int})
+ E
_{H}[\rho^{(1)}_{int},\rho^{(2)}_{int}]
\nonumber\\
&+& E
_{C}[\rho^{(1)}_{int},\rho^{(2)}_{int}]
+ E
_{\Delta kin}[\rho^{(1)}_{int},\rho^{(2)}_{int}]
,
\label{eq:action4}
\end{eqnarray}
where
\footnote{
To obtain these results, we pose $\varphi^{(i)}_{int}=\sqrt{\rho^{(i)}_{int}}$ and use
$
\gamma^{(12)}_{int}({r},{r'})
= \int d{r}^{(1)} d{r}^{(2)} \delta({R}) |\psi_{int}({r}^{(1)}-{r}^{(2)})|^2 \times\\
\delta \big({r}-({r}^{(1)}-{R})\big) \delta \big({r'}-({r}^{(2)}-{R}) \big)
= \frac{1}{2}\Big(\rho^{(1)}_{int}(r)+\rho^{(2)}_{int}(r)\Big) \delta(r+r')
$.
}
\begin{eqnarray}
\label{eq:action5}
&& E
_{H}[\rho^{(1)}_{int},\rho^{(2)}_{int}]
\\
&&\quad= \int d{r} d{r'} \rho^{(1)}_{int}({r}) \rho^{(2)}_{int}({r'}) \frac{1}{4}m\omega^2({r}-{r'})^2
\nonumber\\
&& E
_{C}[\rho^{(1)}_{int},\rho^{(2)}_{int}] 
\label{eq:E12C}\\
&&\quad= 
\int \! d{r} \, d{r'} \gamma^{(12)}_{int}({r},{r'}) \frac{1}{4}m\omega^2({r}-{r'})^2
-E
_{H}[\rho^{(1)}_{int},\rho^{(2)}_{int}]
\nonumber\\
&&\quad= \int d{r} \frac{1}{2}\Big(\rho^{(1)}_{int}(r)+\rho^{(2)}_{int}(r)\Big) m\omega^2{r}^2
-
E
_{H}[\rho^{(1)}_{int},\rho^{(2)}_{int}]
\nonumber\\
&& E
_{\Delta kin}[\rho^{(1)}_{int},\rho^{(2)}_{int}]
\label{eq:E12}\\
&&\quad= (\psi_{int}| \frac{\tau^2}{2\mu} |\psi_{int}) - \sum_{i=1}^2 (\varphi^{(i)}_{int}|\frac{{p}^2}{2m}|\varphi^{(i)}_{int})
\nonumber\\
&&\quad= -\frac{3}{2}\hbar\omega 
+ \frac{3}{2}\int d{r} \Big(\rho^{(1)}_{int}(r)+\rho^{(2)}_{int}(r)\Big) m \omega^2{r}^2
\nonumber
.
\end{eqnarray}
$E_{H}$ is the H energy; $E_{C}$ is the ``standard" correlations energy linked to the particle-particle interaction;
and $E_{\Delta kin}$ is the energy associated to the correlations contained in the interacting kinetic energy.
It is the only term that contains explicitly the c.m. correlations
%
%
\footnote{
Note that, even if in the general case the functional $E_{C}+E_{\Delta kin}$
is universal \cite{Mes09-1,Mes09-2}, the forms (\ref{eq:E12C}) and (\ref{eq:E12}) are limited to the two different particles case
because exchange effects are not taken into account.
Thus they cannot be used to describe a system composed of an arbitrary number of particles of each kind.
The universal functional, applicable to an arbitrary number of particles, is more involved
but should permit us to recover (\ref{eq:E12C}) and (\ref{eq:E12}) to the limit of a system composed by two different particles.
}.

\subsubsection{The c.m. correlations functional.}
\label{app:approxDFT}

We now approximate $E_{\Delta kin}$ by the functional $E_{cm}$
proposed in \S \ref{sec:univ},
with
$
\Gamma^{aux}(R)
=
\Big( \frac{K}{\pi} \Big)^{1/4} \exp\Big\{-\frac{K}{2}R^2\Big\}
$.
%
%
We obtain:
\begin{eqnarray}
\lefteqn{ E
_{cm}
= }
\label{eq:xxx}\\
&&-\frac{\hbar^2}{2m}\int dr 
\Big[
\varphi^{(1)*}_{int}(r)\Delta_r\varphi^{(1)}_{int}(r) \Big( 2 \sqrt{\frac{\pi}{K}}|\varphi^{(2)}_{int}(-r)|^2 -1 \Big) 
\nonumber\\
&&
\quad\quad\quad\quad\quad 
+
\varphi^{(2)*}_{int}(r)\Delta_r\varphi^{(2)}_{int}(r) \Big( 2 \sqrt{\frac{\pi}{K}}|\varphi^{(1)}_{int}(-r)|^2 -1 \Big) 
\Big]
\nonumber\\
&&
-\frac{\hbar^2}{2m} \sqrt{K\pi} \int dr |\varphi^{(1)}_{int}(r)|^2 |\varphi^{(2)}_{int}(-r)|^2
\nonumber
.
\end{eqnarray}
The corresponding local c.m. correlations potentials are ($l=1,2$):
%
\begin{eqnarray}
\label{eq:full}
U_{cm}^{(l)}(r)\varphi^{(l)}_{int}(r) 
&=&
-\frac{\hbar^2}{2m}
\Big( 2 \sqrt{\frac{\pi}{K}}|\varphi^{(m\ne l)}_{int}(-r)|^2 -1 \Big) \Delta_r\varphi^{(l)}_{int}(r)
\nonumber\\
&&
-\frac{\hbar^2}{2m} 
\Big(
2 \sqrt{\frac{\pi}{K}} \varphi^{(m\ne l)*}_{int}(-r)\Delta_r\varphi^{(m\ne l)}_{int}(-r)
\nonumber\\
&&
\quad\quad\quad
+  \sqrt{K\pi} |\varphi^{(m\ne l)}_{int}(-r)|^2 
\Big)
.
\end{eqnarray}
(There is no need of the Slater approximation because we deal with one particle only of each kind.)

Remember that, making this approximation, we are neglecting the ``standard" correlations part of $E_{\Delta kin}$.
In realistic 3D cases these correlations are mostly taken into account in the parametrized functionals that are commonly used.
In our case, there is no simple way to include them in the rest of the functional, so it will not be possible to perfectly match the benchmark.
Nevertheless, as they remain only a correction, the benchmark should be reasonably matched,
at least much better than with the commonly used $-<\frac{\vec{P}^2}{2mN}>$ approximation
(see discussion of \S\ref{sec:current_cm}).

\subsubsection{Numerical results.}
\label{app:results1}

\begin{figure}[ht]
\begin{center}
\includegraphics[angle=-90,width=1.05\linewidth]{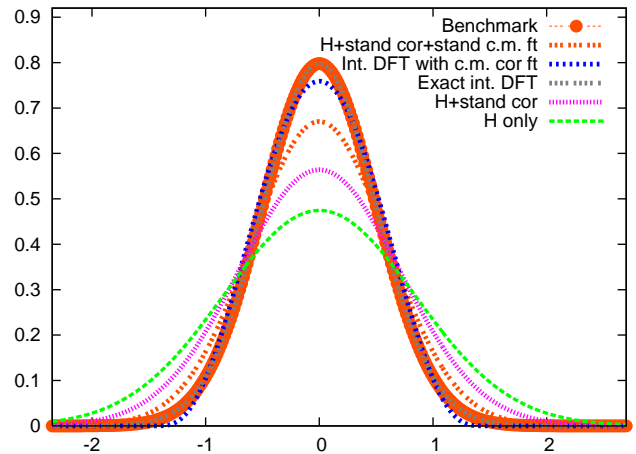}
\caption{
(Color online.)
The internal densities $\rho_{int}^{(l)}$ of the various formalisms
(x-axis: position in units where $\hbar=m=1$).
\label{fig:densities}}
\end{center}
\end{figure}

We use a unit system where $\hbar=m=1$ and choose $\omega=1$.
Table \ref{tab:energies} and Fig.~\ref{fig:densities} give numerical results for the following formalisms:
%
\begin{itemize}
\item $E_H$, called ``H only"
\item $E_H$ $+$ $E_{C}$,
called ``H $+$ standard correlations"
\item $E_H$ $+$ $E_{C}$ $-$ $<\frac{\vec{P}^2}{2mN}>$,
called ``H $+$ standard correlations $+$ standard c.m. correction"
\item $E_H$ $+$ $E_{C}$ $+$ $E^{}_{cm}$, called ``internal DFT with c.m. correlations functional";
we obtain $K\approx 3.9$ with the method described in \S \ref{sec:Gamma_aux}
\item $E_H$ $+$ $E_{C}$ $+$ $E_{\Delta kin}$, called ``exact internal DFT"
\item benchmark (described in \S \ref{app:model})
\end{itemize}
%
%
%
%

First of all, we see from Table \ref{tab:energies} and Fig. \ref{fig:densities} that ``exact internal DFT" perfectly reproduces the total energy and
densities of the benchmark
so that the non-interacting v-representability \cite{Dre90,Mes09-1} is perfectly achieved.
(This is not a surprise; when one deals with only one particle of each kind,
it is always possible to reach $\varphi^{(l)}_{int}=\sqrt{\rho_{int}^{(l)}}$.)

\begin{widetext}
\begin{table}[htbp]
\begin{center}
\begin{tabular}{|l||c|c|c|c||c|}
\hline
\rule[-5pt]{0pt}{20pt}
Formalism&
Non-interacting kin. energy & $E_{H}$ & $E_{C}$ & $-<\frac{\vec{P}^2}{2mN}>$ or $E_{\Delta kin}$ or $E^{}_{cm}$& Total energy\\ 
\hline
\rule[-5pt]{0pt}{20pt}
H only & 0.353 & 0.353 & 0 & 0 & 0.71\\
\hline
\rule[-5pt]{0pt}{20pt}
H $+$ stand. corr. & 0.5 & 0.25 & 0.25 & 0 & 1.00\\ 
\hline
\rule[-5pt]{0pt}{20pt}
H $+$ stand. corr. $+$ stand. c.m. correct. & 0.706 & 0.177 & 0.177 & -0.353 & 0.71\\ 
\hline
\rule[-5pt]{0pt}{20pt}
Internal DFT with c.m. corr. ft & 1.225 & 0.120 & 0.120 & -0.918 & 0.55\\ 
\hline
\rule[-5pt]{0pt}{20pt}
Exact internal DFT & 1.000 & 0.125 & 0.125 & -0.750 & 0.50\\ 
\hline
\end{tabular}
\end{center}
\caption{\label{tab:energies}
The energies of the various formalisms (in units where $\hbar=m=1$; benchmark: total energy $=0.50$;
and interacting kinetic energy $(\psi_{int}|\frac{\tau^2}{2\mu}|\psi_{int})=0.25$).
}
\end{table}
\end{widetext}

From Table \ref{tab:energies}, we see that
the non-interacting kinetic energy 
{cannot} be compared to the interacting kinetic energy.
In particular, there is a factor $4$ between the ``exact internal DFT" non-interacting kinetic energy (equal to $1$)
and the interacting kinetic energy (equal to $0.250$).
Indeed, as discussed in \S \ref{sec:intDFTrecall}, it is the ``non-interacting kinetic energy $+$ $E^{}_{\Delta kin}$"
that is comparable to the interacting kinetic energy.
Exact internal DFT then perfectly reaches the benchmark: $1.000 - 0.750 = 0.250$.
Internal DFT with $E^{}_{cm}$ gives $1.225-0.928=0.307$, which fairly well reproduces the benchmark,
considering that the ``standard" correlations part of $E_{\Delta kin}$ has been neglected.
The result with standard c.m. correction gives $0.706-0.353=0.353$, which is worse.

From the point of view of the total energy, 
``internal DFT with c.m. corr. ft." is much closer to the benchmark than the other approximate schemes.
%
From the point of view of the densities, Fig. \ref{fig:densities} shows that ``internal DFT with c.m. corr. ft." is very close to the benchmark
and represents a great improvement as compared to the other results.
Nevertheless, we see some differences
that are explained by the fact that the ``standard" correlations part of $E_{\Delta kin}$ is not taken into account in our model.

\subsection{Model system of identical Bosons (smooth interaction).}
\label{app:section_smooth}


We now consider a 1D system composed of $N$ identical Bosons of mass $m$ and positions $\{r_i\}$ without spin
in a condensate state and
with an attractive two-body interaction of the form ($e>0$):
\begin{equation}
\label{eq:1Dint}
u(r-r')=-\frac{1}{\sqrt{(r-r')^2+e}}
,
\end{equation}
where the greater $e$, the smoother the potential.
This allows us to model features of $^4$He droplets.

The internal DFT energy functional is given by
($\varphi_{int}$ is the one-body orbital describing the Bosons and $\rho_{int}=N |\varphi_{int}|^2$):
%
\begin{eqnarray}
\label{eq:Eint_bosons}
E_{int}[\rho_{int}] 
& = & N (\varphi_{int}|\frac{{p}^2}{2m}|\varphi_{int}) 
      + E_{H}[\rho_{int}] \times (1-\frac{1}{N})
\nonumber\\
&&    + E_{C}[\rho_{int}] + E_{cm}[\rho_{int}]
,
\end{eqnarray}
where $E_{H}[\rho_{int}] \times (1-\frac{1}{N})$ represents the H energy where the self-interaction has been subtracted 
and $E_{C}[\rho_{int}]$ is the ``standard" correlations energy.
We once again neglect $E_{C}$ because
we have no simple way to evaluate it as a functional of $\rho_{int}$.
Note that more is neglected than in the previous model (of \S\ref{app:model0}),
because $E_{C}$ contains {all} the standard correlations
whereas in \S\ref{app:model0} we were able to keep a part of them.
We thus can expect that the benchmark will be a little less matched here than in \S\ref{app:model0};
nevertheless, as $E_{C}$ is small (even if not always completely negligible), the benchmark should remain reasonably matched.

%
%
The c.m. correlations energy is defined as in \S \ref{sec:bosons},
where $\Gamma^{aux}$ is defined as in \S \ref{sec:Gamma_aux}:
\begin{widetext}
\begin{eqnarray}
E^{}_{cm}[\rho_{int}] &=& 
\label{eq:E12_4-Bosons}
-\frac{\hbar^2}{2m}
\int d{r} \hspace{1mm}
\sqrt{\rho_{int}({r})}\Delta_{\vec{r}}\sqrt{\rho_{int}({r})}
\times
\hspace{1mm} \Big( \sqrt{\frac{\pi}{K(N)}} \int d{r}' \hspace{1mm}
\frac{1}{N} \rho_{int}({r}')
f_2
({r}+{r}') - 1 \Big)
\\
&&
-\frac{\hbar^2}{2mN} 
\sqrt{\pi K(N)}
\int d{r} \hspace{1mm}
\frac{1}{N} \rho_{int}({r})
\int d{r}' \hspace{1mm}
\frac{1}{N} \rho_{int}({r}')
f_2
({r}+{r}')
,
\nonumber
\end{eqnarray}
and
$f_2$ is defined by Eq.~(\ref{eq:f_LDA_2}).
%
%
The internal KS equation is:
\begin{equation}
\Big(
- \frac{\hbar^2}{2m}\Delta 
+ U_H[\rho_{int}]  \times (1-\frac{1}{N})
+ U_{cm}[\rho_{int}]
\Big) \varphi_{int} = \epsilon \varphi_{int}
,
\nonumber
\end{equation}
where:
%
\begin{eqnarray}
U_{cm}[\rho_{int}](r)
&=&
-\frac{\hbar^2}{2m}
\Big\{
\frac{1}{\sqrt{\rho_{int}({r})}} \Delta_{{r}}\sqrt{\rho_{int}({r})}
\times \Big( \sqrt{\frac{\pi}{K(N)}} \int d{r}'
\frac{1}{N} \rho_{int}({r}')
f_2
({r}+{r}') -1 \Big)
\nonumber\\
&&
\quad\quad\quad
+
\sqrt{\frac{\pi}{K(N)}} \hspace{1mm} 
\int d{r}' \hspace{1mm}
\sqrt{\rho_{int}({r}')}\Delta_{\vec{r}}\sqrt{\rho_{int}({r}')}\times
\frac{N-1}{N} 
f_2
({r}+{r}')
\Big\}
\nonumber\\
&&
-\frac{\hbar^2}{2mN} \sqrt{K(N) \pi}
\int d{r}'
\frac{1}{N} \rho_{int}({r}')
\times f_2
({r}+{r}')
.
\label{eq:cm_pot_2_Bosons}
\end{eqnarray}
%
%
\begin{table}[htbp]
\begin{center}
\begin{tabular}{|l||c|c|c||c|}
\hline
\rule[-5pt]{0pt}{20pt}
Formalism&
Non-interacting kin. energy & $E_{H} \times (1-\frac{1}{N})$ & $-<\frac{\vec{P}^2}{2mN}>$ or $E_{cm}$ & Total energy\\ 
\hline
\hline
\rule[-5pt]{0pt}{20pt}
$H$ only & 0.133 & -0.626 & 0 & -0.49\\
\hline
\rule[-5pt]{0pt}{20pt}
H $+$ stand. c.m. correct. & 0.260 & -0.712 & -0.065 & -0.52\\ 
\hline
\rule[-5pt]{0pt}{20pt}
Internal DFT with c.m. corr. ft & 0.535 & -0.776 & -0.418 & -0.66\\ 
\hline
\end{tabular}
\end{center}
\caption{\label{tab:energies_smooth1}
The energies of the various formalisms in the $N=2$ case (in units where $\hbar=m=1$;
benchmark: total energy $=-0.59$;
and interacting kinetic energy $(\psi_{int}|\frac{\tau^2}{2\mu}|\psi_{int})=0.12$).
}
\end{table}
\begin{table}[htbp]
\begin{center}
\begin{tabular}{|l||c|c|c||c||c|}
\hline
\rule[-5pt]{0pt}{20pt}
$N$ & Non-interacting kin. energy & $E_{cm}$ & Total energy  & $K$ & Interacting kin. energy\\ 
\hline
\hline
\rule[-5pt]{0pt}{20pt}
2 & 0.535 & -0.418 & -0.66 & 1.94 & 0.117\\
\hline
\rule[-5pt]{0pt}{20pt}
3 & 0.463 & -0.185 & -1.90 & 1.74 & 0.278\\ 
\hline
\rule[-5pt]{0pt}{20pt}
4 & 0.702 & -0.196 & -3.97 & 2.74 & 0.507\\ 
\hline
\rule[-5pt]{0pt}{20pt}
5 & 1.014 & -0.217 & -6.84 & 4.01 & 0.799\\ 
\hline
\rule[-5pt]{0pt}{20pt}
6 & 1.390 & -0.239 & -10.56 & 4.54 & 1.151\\ 
\hline
\end{tabular}
\end{center}
\caption{\label{tab:energies_smooth_DFT}
The ``internal DFT with c.m. corr. ft." energies for various $N$ (in units where $\hbar=m=1$).
}
\end{table}
\end{widetext}

\begin{table}[htbp]
\begin{center}
\begin{tabular}{|l||
c|}
\hline
\rule[-5pt]{0pt}{20pt}
$N$ & 
Total energy\\ 
\hline
\hline
\rule[-5pt]{0pt}{20pt}
2 & 
-0.491\\
\hline
\rule[-5pt]{0pt}{20pt}
3 & 
-1.75\\ 
\hline
\rule[-5pt]{0pt}{20pt}
4 & 
-3.79\\ 
\hline
\rule[-5pt]{0pt}{20pt}
5 & 
-6.66\\ 
\hline
\rule[-5pt]{0pt}{20pt}
6 & 
-10.37\\ 
\hline
\end{tabular}
\end{center}
\caption{\label{tab:energies_smooth_HF}
The ``H only" energies for various $N$ (in units where $\hbar=m=1$).
}
\end{table}

\begin{figure}[ht]
\begin{center}
\includegraphics[angle=-90,width=1.05\linewidth]{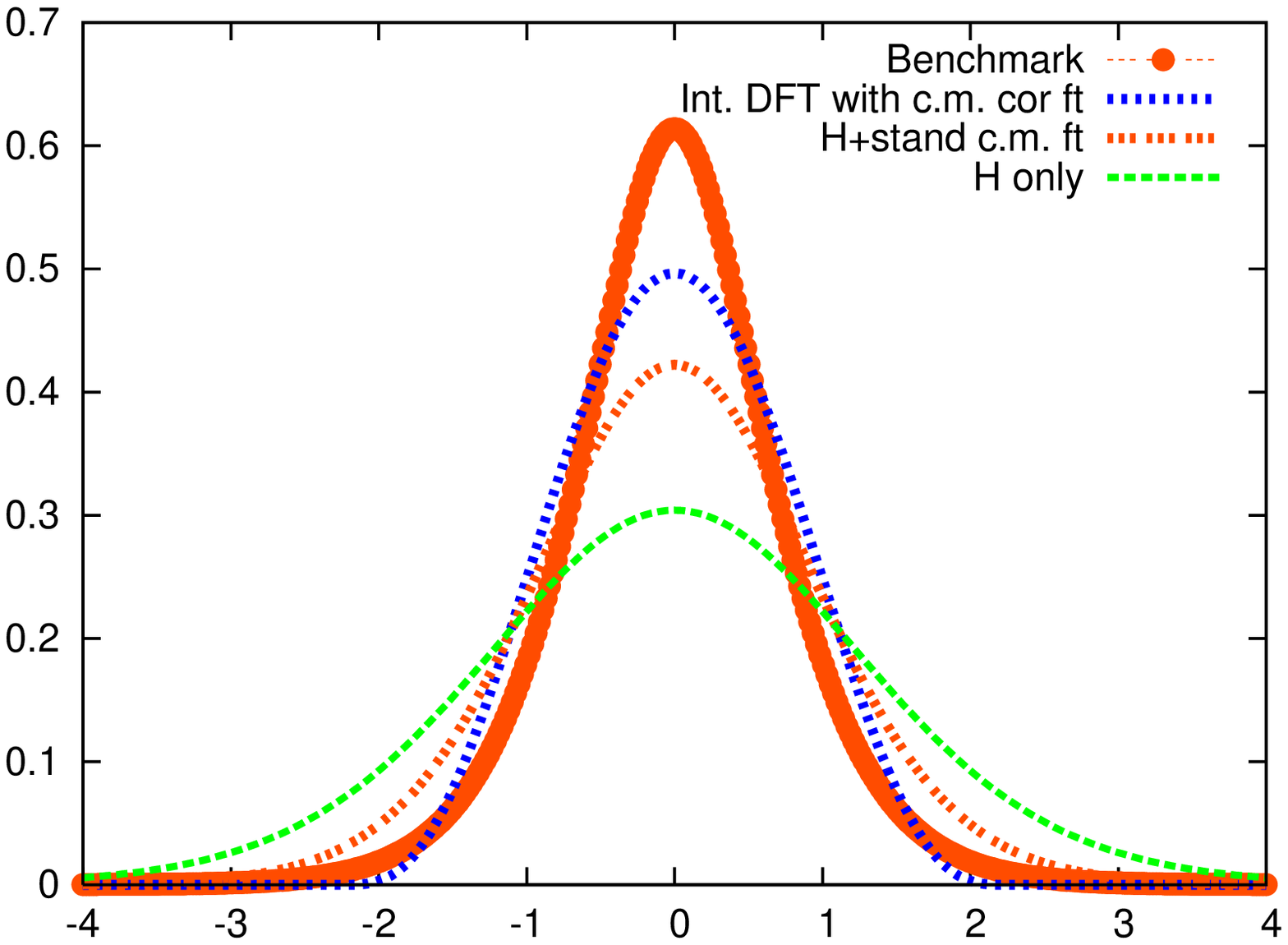}
\caption{
(Color online.)
The internal density $\rho_{int}/2$ of the various formalisms in the $N=2$ case
(x-axis: position in units where $\hbar=m=1$).
\label{fig:densities_smooth1}}
\end{center}
\end{figure}
\begin{figure}[ht]
\begin{center}
\includegraphics[angle=-90,width=1.05\linewidth]{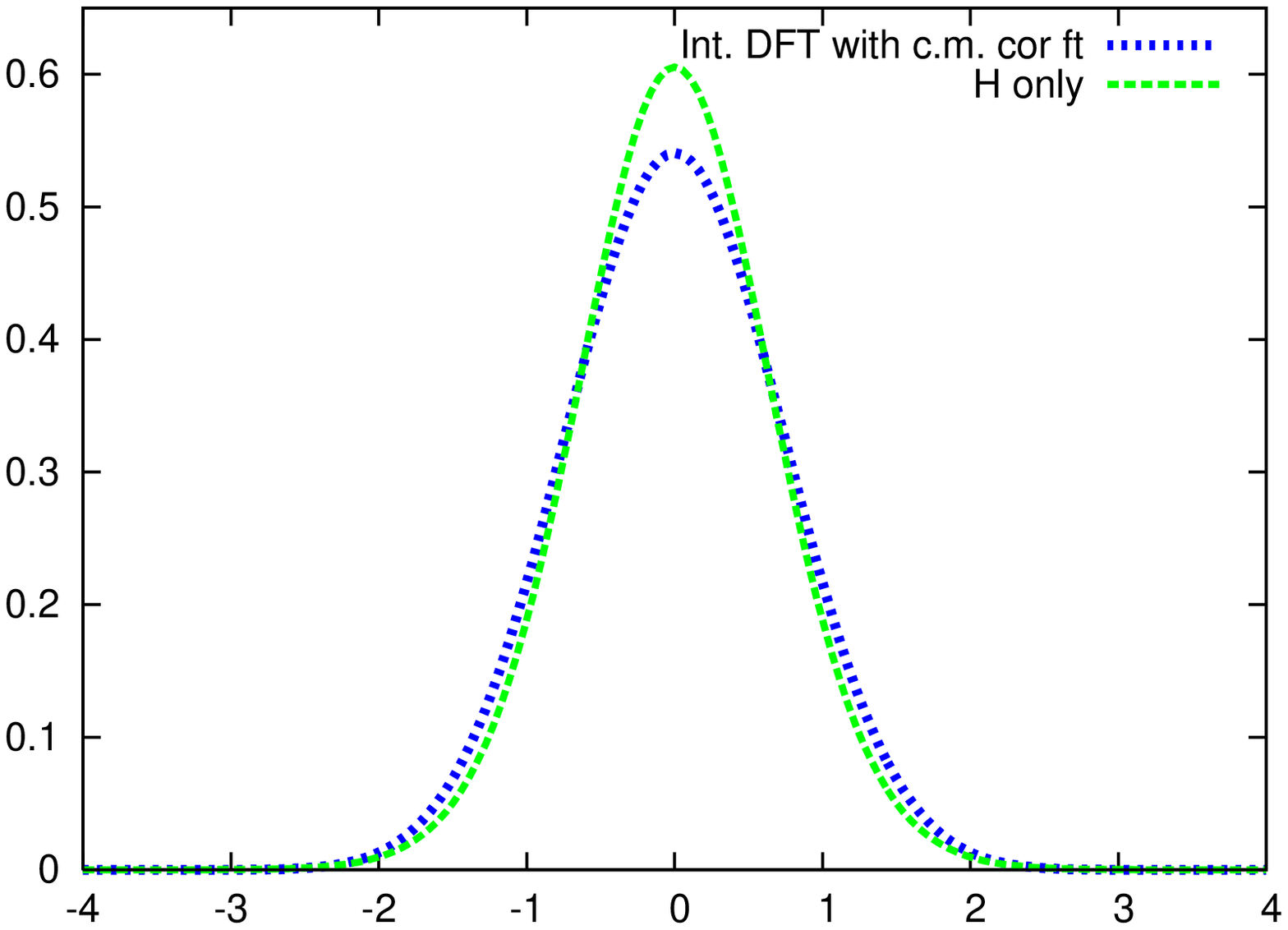}
\caption{
(Color online.)
The internal density $\rho_{int}/6$ of the various formalisms in the $N=6$ case
(x-axis: position in units where $\hbar=m=1$).
\label{fig:densities_smooth2}}
\end{center}
\end{figure}
\begin{figure}[ht]
\begin{center}
\includegraphics[angle=-90,width=1.05\linewidth]{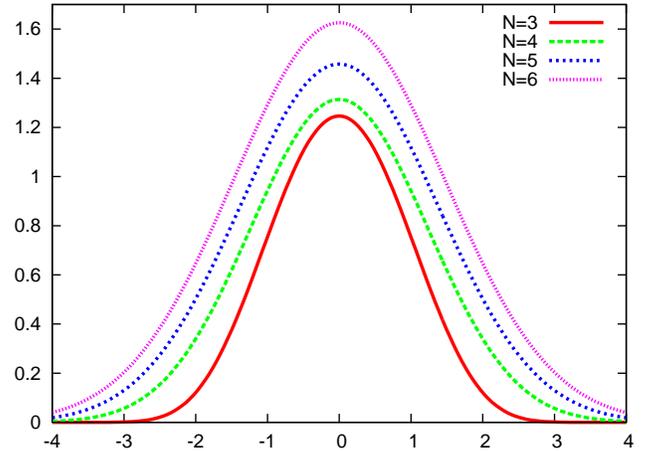}
\caption{
(Color online.)
$f_2$ for $N=$ $3$ to $6$ ($N=2$ is the delta function; x-axis: position in units where $\hbar=m=1$).
\label{fig:densities_smooth3}}
\end{center}
\end{figure}

For the $N=2$ case, we can compute a benchmark.
Indeed, by use of Jacobi coordinates, the internal Hamiltonian can be rewritten as
$H_{int}=\frac{\tau^2}{2\mu} - \frac{1}{\sqrt{\xi^2+e}}$,
where $\mu=m/2$ is the reduced mass.
It is then possible to calculate numerically the exact many body ground state $\psi_{int}$
and the c.m. frame one-body density $\rho^{}_{int}(r)=4 |\psi_{int}(2 r)|^2$.

The next results are given for:
\begin{itemize}
\item $E_H\times (1-\frac{1}{N})$, called ``$H$ only"
\item $E_H\times (1-\frac{1}{N})$ $-$ $<\frac{\vec{P}^2}{2mN}>$, called ``H $+$ standard c.m. correction"
\item $E_H\times (1-\frac{1}{N})$ $+$ $E^{}_{cm}$, called ``internal DFT with c.m. correlations functional"
\item benchmark (for the $N=2$ case only)
\end{itemize}

Table \ref{tab:energies_smooth1} and Fig.~\ref{fig:densities_smooth1}
show the energies and densities for the $N=2$ case.
We again see that the internal DFT non-interacting kinetic energy {cannot} be compared to the interacting kinetic energy.
It is the internal DFT ``non-interacting kinetic energy $+$ $E^{}_{cm}$" ($0.535-0.418=0.117$) that is comparable to the interacting kinetic energy ($0.12$).
The result with standard c.m. correction gives $0.260-0.065=0.195$, which is worse.
We also see that ``internal DFT with c.m. corr. ft."
reproduces fairly well the benchmark, at least much better than the other schemes.
We nevertheless see some differences, which are explained by the fact that $E_{C}$ has been neglected.
(See discussion at the beginning of this section.)

Table \ref{tab:energies_smooth_DFT} shows that the part of the c.m. correlations energy in the total internal DFT energy decreases
as $N$ grows ($63\%$ for $N=2$; $2\%$ for $N=6$).
As a consequence,
and even if shell effets play a role,
the internal DFT non-interacting kinetic energy tends to become closer to the interacting kinetic energy
as $N$ grows (factor $4.6$ for $N=2$; factor $1.2$ for $N=6$).
The ``H only" and internal DFT densities become closer as N grows (see Fig.~\ref{fig:densities_smooth2} for $N=6$).
Tables \ref{tab:energies_smooth_DFT} and \ref{tab:energies_smooth_HF}
show that the ``H only" total energy also becomes closer to internal DFT total energy as N grows.

Finally, we see from Table \ref{tab:energies_smooth_DFT} that $K$ grows as $N$ grows
and from Fig.~\ref{fig:densities_smooth3} that the maximum value of $f_2$ also grows as $N$ grows
(but the maximum value $\frac{1}{N^D}f_2$ diminishes),
confirming the reasoning of \S \ref{sec:Gamma_aux}.

\section{Conclusion.}

\JMcomm{
Internal DFT provides an existence theorem for a c.m. correlations energy functional
associated to a local potential.
In this article, we have constructed such a functional,
without involving any free parameters.
The use of this functional is justified by a strong formal background
and variants suitable for Fermionic as well as Bosonic systems have been proposed.
The resulting scheme is numerically manageable
and represents a well-founded alternative to projection techniques to treat the c.m. correlations.
It can directly be added to actual energy functionals
although a refitting of them then would be necessary.
Moreover, this scheme permits us to recover the precise value of the interacting kinetic energy and
represents a manageable way to include
the c.m. correlations in time-dependent calculations
of self-bound systems.
}

\JMcomm{
We have presented convincing numerical results on 1D model systems.
These results show that the developed functional
represents a great improvement compared to the ``standard c.m. correction" commonly used in nuclear physics
(of the form $-<\frac{\vec{P}^2}{2mN}>$), especially from the point of view of the energies.
The next step will be to include the proposed functional in
realistic 3D calculations, for instance in mean-field-like calculations of nuclei with Skyrme interaction~\cite{RS80,Ben03}.
As the ``standard" correlations are mostly taken into account in the commonly used functionals,
the 3D results should be even more convincing than the 1D ones.
}

\JMcomm{
Even if the proposed Gaussian set for $\Gamma^{aux}$ has been proved to give satisfying results,
the search for other forms,
i.e., with other variations around $\vec{R}=\vec{0}$, should continue
to provide the most precise description of atomic nuclei,
helium droplets or small molecular systems where a quantum treatment of the nuclei is necessary.
}

\subsection*{Acknowledgments.}

The author is particularly grateful to T. Duguet and J. Donohue for numerous enlightening discussions and
reading of the manuscript,
and to S. Bernard for careful reading of the manuscript.
The author also gratefully acknowledges the Conseil R\'egional d'Aquitaine for support.\\
\\


\begin{appendix}

\section{Details of the calculation that leads to $E_{cm}$.}
\label{app:app1}

We evaluate 
$\int d{\vec{r}}_1\dots d{\vec{r}}_N \delta({\vec{R}}) 
\psi^{*}_{int}({\vec{r}}_1,\dots,{\vec{r}}_N)\\
\times\sum_{i=1}^N \frac{\vec{p}_i^2}{2m}
\psi_{int}({\vec{r}}_1,\dots,{\vec{r}}_N)
$
using the approximation (\ref{eq:zzz2}) for $\psi_{int}$ and obtain:
%
%
%
\begin{widetext}
\begin{eqnarray}
\label{eq:E12_3}
\int && d{\vec{r}}_1\dots d{\vec{r}}_N \delta({\vec{R}}) 
\frac{1}{\Gamma^{aux*}(\vec{R})} \psi^{aux*}({\vec{r}}_1,\dots,{\vec{r}}_N)
\sum_{i=1}^N \frac{\vec{p}_i^2}{2m}
\frac{1}{\Gamma^{aux}(\vec{R})} \psi^{aux}({\vec{r}}_1,\dots,{\vec{r}}_N)
\\
&=&
-\frac{\hbar^2}{2m}\frac{1}{|\Gamma^{aux}(\vec{0})|^2}
\frac{1}{N!}\sum_{P,P'} (-1)^{p+p'}
\sum_{i=1}^N \int d\vec{r}_i  \hspace{1mm}
F_i^{P,P'}[\{\varphi^k_{int}\}](\vec{r}_i)
\times
\varphi^{P(i)*}_{int}(\vec{r}_i)\Delta_{\vec{r}_i}\varphi^{P'(i)}_{int}(\vec{r}_i)
\nonumber\\
&&
-\frac{\hbar^2}{2m} \frac{1}{N^2} \frac{1}{\Gamma^{aux*}(\vec{0})} \Delta_\vec{R} \frac{1}{\Gamma^{aux}(\vec{R})}\Big|_{\vec{R}=\vec{0}}
\frac{1}{N!}\sum_{P,P'}(-1)^{p+p'}
\sum_{i=1}^N \int d\vec{r}_i \hspace{1mm}
F_i^{P,P'}[\{\varphi^k_{int}\}](\vec{r}_i)
\times
\varphi^{P(i)*}_{int}(\vec{r}_i)\varphi^{P'(i)}_{int}(\vec{r}_i)
\nonumber\\
&=&
-\frac{\hbar^2}{2m}\frac{1}{|\Gamma^{aux}(\vec{0})|^2}
\sum_{i=1}^N \Big\{
\int d\vec{r} \hspace{1mm}
F_i[\{\varphi^{k\ne i}_{int}\}](\vec{r})
\times
\varphi^{i*}_{int}(\vec{r})\Delta_{\vec{r}}\varphi^{i}_{int}(\vec{r})
\nonumber\\
&&
\hspace{3.3cm}
+\frac{1}{N!}\sum_{P,P'\ne P}(-1)^{p+p'} \int d\vec{r} \hspace{1mm}
F_i^{P,P'\ne P}[\{\varphi^{k}_{int}\}](\vec{r})
\times
\varphi^{P(i)*}_{int}(\vec{r})\Delta_{\vec{r}}\varphi^{P'(i)}_{int}(\vec{r}) 
\Big\}
\nonumber\\
&&
-\frac{\hbar^2}{2mN} \frac{1}{\Gamma^{aux*}(\vec{0})} \Delta_\vec{R} \frac{1}{\Gamma^{aux}(\vec{R})}\Big|_{\vec{R}=\vec{0}} \frac{1}{N} 
\sum_{i=1}^N\Big\{
\int d\vec{r} \hspace{1mm}
F_i [\{\varphi^{k\ne i}_{int}\}](\vec{r})
\times
|\varphi^{i}_{int}(\vec{r})|^2
\nonumber\\
&&
\hspace{6.5cm}
+\frac{1}{N!}\sum_{P,P'\ne P}(-1)^{p+p'} \int d\vec{r} \hspace{1mm}
F_i^{P,P'\ne P}[\{\varphi^{k}_{int}\}](\vec{r})
\times
\varphi^{P(i)*}_{int}(\vec{r})\varphi^{P'(i)}_{int}(\vec{r}) 
\Big\}
\nonumber
,
\end{eqnarray}
\end{widetext}
where we have defined ($D$ = 1, 2 or 3 is the dimension in which the calculation is done):
\begin{eqnarray}
\lefteqn{ F_i^{P,P'}[\{\varphi^k_{int}\}](\vec{r})=}
\nonumber\\ 
&& N^D\int \Pi_{\stackrel{j=1}{j\ne i}}^N d\vec{r}_j \delta\big(\sum_{\stackrel{k=1}{k\ne i}}^N \vec{r}_k + \vec{r}\big) \Pi_{\stackrel{j=1}{j\ne i}}^N \varphi^{P(j)*}_{int}(\vec{r}_j) \varphi^{P'(j)}_{int}(\vec{r}_j)
,
\nonumber
\end{eqnarray}
and its diagonal part:
%
\begin{eqnarray}
\label{eq:Fi}
\lefteqn{ F_i[\{\varphi^{k\ne i}_{int}\}](\vec{r})=\frac{1}{N!} \sum_P F_i^{P,P}[\{\varphi^k_{int}\}](\vec{r}) =}
\nonumber\\
&& N^D \int \Pi_{\stackrel{j=1}{j\ne i}}^N d\vec{r}_j \delta\big(\sum_{\stackrel{k=1}{k\ne i}}^N \vec{r}_k + \vec{r}\big) \Pi_{\stackrel{j=1}{j\ne i}}^N |\varphi^{j}_{int}(\vec{r}_j)|^2
.
\end{eqnarray}
$F_i$ 
is the probability that particle $i$ has position $\vec{r}$, according to the c.m. coupling with every other particles and their probability distributions.
In the following, we will note $F_i$ instead of $F_i[\{\varphi^{k\ne i}_{int}\}]$
to lighten the notations.

$F_{P,P'\ne P}$ is only due to exchange effects.
In all the following, as explained in \S\ref{sec:method_proposed} and \S\ref{sec:The functional},
we neglect the pure exchange effects and thus $F_{P,P'\ne P}$.
We obtain:
%
\begin{widetext}
\begin{eqnarray}
\label{eq:E12_4}
&&\int d{\vec{r}}_1\dots d{\vec{r}}_N \delta({\vec{R}}) 
\psi^{*}_{int}({\vec{r}}_1,\dots,{\vec{r}}_N)
\sum_{i=1}^N \frac{{p}_i^2}{2m}
\psi_{int}({\vec{r}}_1,\dots,{\vec{r}}_N)
\quad\rightarrow
\nonumber\\
&&
-\frac{\hbar^2}{2m}\frac{1}{|\Gamma^{aux}(\vec{0})|^2}
\sum_{i=1}^N \int d\vec{r} \hspace{1mm}
F_i(\vec{r})
\times
\varphi^{i*}_{int}(\vec{r})\Delta_{\vec{r}}\varphi^{i}_{int}(\vec{r})
-\frac{\hbar^2}{2mN} \frac{1}{\Gamma^{aux*}(\vec{0})} \Delta_\vec{R} \frac{1}{\Gamma^{aux}(\vec{R})}\Big|_{\vec{R}=\vec{0}}
\times \frac{1}{N} \sum_{i=1}^N \int d\vec{r} \hspace{1mm}
F_i(\vec{r})
\times
|\varphi^{i}_{int}(\vec{r})|^2
\nonumber
.
\end{eqnarray}
We now insert this result in $E^{}_{\Delta kin}$, Eq.~(\ref{eq:Exc__2}),
and keep only the real part, i.e., $\Re e (E^{}_{\Delta kin})$, as justified in \S \ref{sec:The functional}.
We are left only with the c.m. correlations contribution:
%
\begin{eqnarray}
E^{}_{\Delta kin} \rightarrow E^{}_{cm}[\{\varphi^{k}_{int}\}] &=&
-\frac{\hbar^2}{2m}
\sum_{i=1}^N \int d\vec{r} \hspace{1mm}
\Big(
\frac{1}{|\Gamma^{aux}(\vec{0})|^2} F_i(\vec{r}) -1
\Big)
\times
\varphi^{i*}_{int}(\vec{r})\Delta_{\vec{r}}\varphi^{i}_{int}(\vec{r})
\label{eq:finalE}
\\
&&
-\frac{\hbar^2}{2mN} 
\frac{1}{\Gamma^{aux*}(\vec{0})} \Delta_\vec{R} \frac{1}{\Gamma^{aux}(\vec{R})}\Big|_{\vec{R}=\vec{0}}
\int d\vec{r} \hspace{1mm}
F_i(\vec{r})
\times
|\varphi^{i}_{int}(\vec{r})|^2
\nonumber\\
&&
\JMcomm{
+ \hspace{1mm}
PureImaginary[\{\varphi^{k}_{int}\}]
}
\nonumber
,
\end{eqnarray}
\end{widetext}
where
$PureImaginary[\{\varphi^{k}_{int}\}]$
is a pure imaginary functional which counteracts the imaginary part of
the first two lines of (\ref{eq:finalE}).

%
$F_i$ is interesting in terms of the physics in energy considerations,
although it is not a fundamental quantity for the potential
(obtained by variation of $E^{}_{cm}$).
We thus introduce a more fundamental quantity
which will appear in both the c.m. correlations energy and potential, 
namely the ``two-particle c.m. correlations functional" defined in Eq.~(\ref{eq:f}),
which is linked to $F_i$ by the relation:
\begin{eqnarray}
\forall l\ne i:
&& F_i(\vec{r}') = \int d\vec{r} |\varphi^{l}_{int}(\vec{r})|^2 
f_{i,l\ne i}(\vec{r}+\vec{r}')
.
\label{eq:Fi_fil}
\end{eqnarray}
When (\ref{eq:Fi_fil}) is inserted in (\ref{eq:finalE}),
we obtain the form (\ref{eq:E12_4-}) for the c.m. correlation energy.

\section{The multiconvolution theorem.}
\label{app:app2}

We define the Fourier transform $\mathcal{T}$ of an integrable function $L:\Re e \rightarrow \Im m$ as:
\begin{eqnarray}
\forall \vec{r}, \vec{s} \in \Re e:\quad
\mathcal{T}[L](\mathbf{s})
= \int d{\vec{r}} \hspace{1mm} e^{ -2\pi i \vec{s}.{\vec{r}} } \hspace{1mm} L({\vec{r}})
\label{eq:FT}
\end{eqnarray}
and the inverse Fourier transform $\mathcal{T}^{-1}$ of a function $\tilde{L}:\Re e \rightarrow \Im m$ as:
$$
\forall \vec{r}, \vec{s} \in \Re e:\quad
\mathcal{T}^{-1}[\tilde{L}](\mathbf{r})
= \int d{\vec{s}} \hspace{1mm} e^{ 2\pi i \vec{s}.{\vec{r}} } \hspace{1mm} \tilde{L}({\vec{s}})
.
$$

We start from ($K+1$) integrable functions $g_i:\Re e\rightarrow \Im m$ and define the ``multiconvolution":
\begin{eqnarray}
\lefteqn{ C[\{g_i\}](\tilde{\vec{r}})=}
\nonumber\\
&& \int d\vec{r}_1 \dots d\vec{r}_K \hspace{1mm} g_1(\vec{r}_1) \times \dots \times g_K(\vec{r}_K) 
\times g_{K+1} (-\sum_{i=1}^K \vec{r}_i - \tilde{\vec{r}})
.
\nonumber
\end{eqnarray}
We can show easilly that:
%
\begin{eqnarray}
\mathcal{T}[C](\mathbf{s})
&=&
\Pi_{i=1}^{K+1} \mathcal{T}[g_i](-\mathbf{s})
,
\end{eqnarray}
%
that leads to:
\begin{eqnarray}
\lefteqn{ C[\{g_i\}](\tilde{\vec{r}}) } 
\nonumber\\
&&= \mathcal{T}^{-1} \Big[ \Pi_{i=1}^{K+1} \mathcal{T}[g_i] (-\mathbf{s}) \Big] (\tilde{\vec{r}})
= \mathcal{T}^{-1} \Big[ \Pi_{i=1}^{K+1} \mathcal{T}[g_i] \Big] (-\tilde{\vec{r}})
.
\nonumber
\end{eqnarray}
This is the ``convolution theorem" \cite{???2} generalized to multiconvolutions
which states that the Fourier transform of a multiconvolution is the product of the Fourier transforms
of each function that enters into the multiconvolution.
%
Note that this relationship is only valid for the form (\ref{eq:FT}) of the Fourier transform.
For forms normalized in other ways, a constant scaling factor
will appear.

\JMcomm{
\section{Some properties of $\Gamma^{aux}$ when $N$ becomes very large.}
\label{app:app3}
}

\JMcomm{
The limit where the c.m. correlations become negligible is obtained
when $N$ becomes very large, as mentioned in \S \ref{sec:c.m. cor functional}. 
Indeed, $f_{i,l\ne i}$ then tends to become constant and delocalized in the whole space.
We define:
\begin{eqnarray}
\lim_{N\rightarrow +\infty} f_{i,l\ne i} = Constant
.
\label{eq:lim_f}
\end{eqnarray}
}
\JMcomm{
The normalization condition (\ref{eq:normal}) thus implies, when $N$ is very large:
\begin{eqnarray}
\lefteqn{ \lim_{N\rightarrow +\infty} |\Gamma^{aux}(\vec{0})|^2 }
\label{eq:lim_gamma}\\
&& = 
\lim_{N\rightarrow +\infty} \int d\vec{r} \hspace{1mm} d\vec{r}' \hspace{1mm}
|\varphi^{i}_{int}(\vec{r})|^2 |\varphi^{l\ne i}_{int}(\vec{r}')|^2 f_{i,l\ne i}(\vec{r}+\vec{r}')
\nonumber\\
&& = 
Constant
.
\nonumber
\end{eqnarray}
%
(Remind that $\Gamma^{aux}$ is implicitly dependent of $N$.)
When these results are inserted in $E_{cm}$, Eq.~(\ref{eq:E12_4-}), we see that its second line becomes null,
and that its third line becomes proportional to 
$\frac{1}{N}\times\frac{1}{\Gamma^{aux*}(\vec{0})} \Delta_\vec{R} \frac{1}{\Gamma^{aux}(\vec{R})}\Big|_{\vec{R}=\vec{0}}\times |\Gamma^{aux}(\vec{0})|^2$
which must tend to zero when $N$ becomes very large so that $E_{cm}$ can be neglected.
This implies the first relation that $\Gamma^{aux}$ should satisfy:
\begin{eqnarray}
\lim_{N\rightarrow +\infty}
\frac{1}{N}\times\frac{1}{\Gamma^{aux*}(\vec{0})} \Delta_\vec{R} \frac{1}{\Gamma^{aux}(\vec{R})}\Big|_{\vec{R}=\vec{0}}\times |\Gamma^{aux}(\vec{0})|^2
\rightarrow 0
.
\nonumber
\end{eqnarray}
}

\JMcomm{
We denote $\cal R$ the region of space where the system has a non-zero density and $V$ the corresponding volume.
For very large systems, we have:
\begin{eqnarray}
|\varphi^i_{int}(\vec{r})|^2 &\approx \frac{1}{V},& \quad
\mbox{ for } {\bf r}\in{\cal R},
\nonumber\\
&\approx 0,& \quad
\mbox{ for } {\bf r}\notin{\cal R}.
\end{eqnarray}
Inserting those results in the definition (\ref{eq:fN>4}) of $f_{i,l\ne i}$ gives:
\begin{eqnarray}
f_{i,l\ne i}({\vec{r}}) &\approx \frac{N^D}{V},& \quad
\mbox{ for } {\bf r}\in{\cal R},
\nonumber\\
&\approx 0,& \quad
\mbox{ for } {\bf r}\notin{\cal R}.
\end{eqnarray}
In the general case, we have $V < kN$, where $k$ is a constant
(as for saturating systems, like nuclear ones \cite{RS80}, where $V$ becomes close, but still inferior, to $k N$).
Thus:
\begin{eqnarray}
\lim_{N\rightarrow +\infty} f_{i,l\ne i}(\tilde{\vec{r}}) =  +\infty
,
\label{eq:lim__f}
\end{eqnarray}
whatever the dimension in which the calculation is done
(but $\lim_{N\rightarrow +\infty} \frac{1}{N^D} f_{i,l\ne i}({\vec{r}})=\lim_{N\rightarrow +\infty} \frac{1}{V}=0$).
As a consequence of Eqs.~(\ref{eq:lim_f}), (\ref{eq:lim_gamma}) and (\ref{eq:lim__f}),
we deduce a second relation that $\Gamma^{aux}$ should satisfy:
\begin{eqnarray}
\lim_{N\rightarrow +\infty} |\Gamma^{aux}(\vec{0})|^2 \rightarrow  +\infty
.
\end{eqnarray}
}


\end{appendix}

\end{document}